\documentclass{aa}

\usepackage{graphicx}
\usepackage{txfonts}

\usepackage{academicons}

\usepackage{natbib,twoopt}
\newcommand{\alfa}{$\alpha$}
\newcommand{\AF}{[\alfa/Fe]}
\newcommand{\Meta}{[M/H]} 
\newcommand{\FeH}{[Fe/H]}
\newcommand{\NdII}{Nd~{\sc ii}}

\newcommand{\SNR}{$S/N$}
\newcommand{\Ms}{M$_\odot$}

\usepackage[dvipsnames]{xcolor}

\newcommand{\Gaia}{{\it Gaia}}
\newcommand{\flags}{{\it flags\_gspspec}}
\newcommand{\gspspec}{{\it GSP-Spec}}
\newcommand{\T}{$T_{\rm eff}$}
\newcommand{\g}{log($g$)}

\newcommand{\meta}{[M/H]}

\newcommand{\AGBNd}{{\it Nd AGB sample}}
\newcommand{\LUSCe}{{\it low-uncertainty Ce sample}}
\newcommand{\AGBCe}{{\it Ce AGB sample}}
\newcommand{\AGBs}{{\it $s$-process AGB sample}}
\newcommand{\AGB}{{\it best-parameterised AGB}}
\newcommand{\alphaFe}{[$\alpha$/Fe]}

\newcommand{\CeFe}{[Ce/Fe]}
\newcommand{\CaFe}{[Ca/Fe]}
\newcommand{\NdFe}{[Nd/Fe]}

\usepackage{listings}
\definecolor{dkgreen}{rgb}{0,0.6,0}
\definecolor{gray}{rgb}{0.5,0.5,0.5}
\definecolor{mauve}{rgb}{0.58,0,0.82}

\begin{document}

   \title{Production of $s$-process elements in AGB stars as revealed by \Gaia/\gspspec\ abundances}


   \author{G. Contursi
          \inst{1}
          \and
          P. de Laverny \inst{1}
          \and
          A. Recio-Blanco\inst{1}
          \and
          P. A. Palicio\inst{1}
         \and
         C. Abia\inst{2}
          }

   \institute{Université Côte d'Azur, Observatoire de la Côte d'Azur, CNRS, Laboratoire Lagrange, Bd de l'Observatoire, CS 34229, 06304 Nice cedex 4, France  
   \and
   Departmento de Fisica Teorica y del Cosmos, Universidad de Granada, E-18071 Granada, Spain
   }

   \date{Received ?? ; accepted ??}
   
   \abstract
   {The recent parameterisation by the \gspspec\ module of \Gaia/Radial Velocity Spectrometer stellar spectra has produced an homogeneous catalogue of about 174,000 Asymptotic Giant Branch (AGB) stars. Among the 13 chemical elements presented in this \Gaia\ third data release, the abundance of two of them (cerium and neodynium) have been estimated in most of these AGB.  These two species are formed by slow neutron captures ($s$-process) in the interior of low- and intermediate-mass stars. They belong to the family of second peak $s$-process  elements.}
   {We study the content and production rate of Ce and Nd in AGB stars, using the atmospheric parameters and chemical abundances derived by the \gspspec~module.}
   {We define a working sample of 19,544 AGB stars having high-quality Ce and/or Nd abundances, selected by applying a specific combination of the \gspspec\ quality flags. We compare these abundances with the yield production predicted by AGB evolutionary models.}
   {We first confirmed that the majority of the working sample is composed of AGB stars by estimating their absolute magnitude in the $K$-band and their properties in a \Gaia-2MASS diagram. We also checked that these stars are oxygen-rich AGBs, as assumed during the \gspspec\ parameterisation. A good correlation between the Ce and Nd abundances is found, confirming the high quality of the derived abundances and that these species indeed belong to the same $s$-process family. We also found higher Ce and Nd abundances for more evolved AGB stars of similar metallicity, illustrating the successive mixing episodes enriching the AGB surface in $s$-process elements formed deeper in their stellar interior. We then compared the observed Ce and Nd abundances with FRUITY and Monash AGB yields and found that the higher Ce and Nd abundances can not be explained by AGBs of mass higher than 5~\Ms. On the contrary, the yields predicted by both models for AGB with an initial mass between $\sim$1.5 and $\sim$2.5~\Ms\ and metallicities between $\sim$-0.5 and $\sim$0.0~dex  are fully compatible with the observed \gspspec\  abundances. 
   }
   {This work, based on the largest catalogue of high-quality second-peak $s$-elements abundances in oxygen-rich AGB, allows to constrain evolutionary models and confirms the fundamental role played by low- and intermediate-mass stars in the enrichment of the Universe in these chemical species.}
   
   
   \maketitle

\section{Introduction}

Asymptotic Giant Branch stars (AGB, hereafter) correspond to the late evolutionary stages of low (masses smaller than $\sim$3~\Ms) and intermediate-mass (between $\sim$4 and $\sim$8~\Ms) stars. Due to their specific internal structure, efficient mixing events and high-mass loss rates, AGBs are among the main contributors to the interstellar medium enrichment in several species. They hence play a fundamental role in the chemical evolution of the Universe \citep{Ulrich73}. Among all the elements produced by AGB, there are neutron capture elements formed through the slow {neutron-capture} process (so-called $s$-process) that are of special interest. Indeed, Solar system abundances distribution of elements formed through the $s$-process show three peaks located around the atomic mass numbers A = 90, 138 and 208 (corresponding to the magic number of neutrons: 50, 82 and 126). While massive stars ($\gtrsim$ 8-10~M$_\odot$) and massive AGBs are the main contributors of first peak $s$-elements \citep{Peters68, Lamb77, Pignatari10, Limongi18}, species of the second peak such as Ce and Nd are mainly formed within AGB stars of lower masses, in which the main neutron source is the $^{13}$C($\alpha$, n)$^{16}$O reaction \citep{Arlandini99, Busso99, KL14, Bisterzo11, Bisterzo15}. Finally, third peak $s$-process elements such as Pb are thought to be predominantly formed within low-mass and low-metallicity AGBs \citep{Gallino98, Choplin22}.


We remind that the internal structure of an AGB stars (mass $\le$ 8 \Ms) is made by a compact and degenerated C-O core, a thin He-burning shell and an H-burning shell separated by a He-intershell \cite[composed of about 75\% of He and  22\% of C according to][]{Karakas02}. All these components are surrounded by a convective H- (and He-) rich envelope which is plagued by large mass-loss rates (from 10$^{-8}$ to 10$^{-4}$ M$_\odot$/yr). During the AGB phase, material formed in the internal layers of the star (such as carbon and $s-$process elements) is brought to the surface by successive penetration of the convective envelope in the He-intershell. This phenomenon is known as the Third Dredge-Up (TDU, hereafter) \citep[see, e.g.][]{Straniero03}. 
During their lifetime, AGB stars experience several Thermal Pulses (TP, hereafter) and hence several mixing episodes (referred as several TDU episodes, hereafter). However, the number of TP strongly depends on the mass loss as well as on the initial mass and metallicity and affects the nucleosynthesis occurring within AGB stars. Actually, theoretical predictions of the number of TP for a given star are still rather uncertain, since the lifetime on the AGB is rather not well known. 

Several stellar evolution models including nucleosynthesis of $s$-process elements have been developed to interpret the observed chemo-physical properties of AGB. The two most complete AGB models published up to now are the FRUITY\citep{Cristallo09, Cristallo11, Cristallo15} and the Monash models \citep{Lugaro12, Fishlock14, Karakas16, Karakas18}.
Both sets of models have their own AGB nucleosynthesis prediction which may differ due to the different physical assumptions adopted during the model computation. The predicted yields are essential for the computation of chemical evolution models, but also for the direct comparison with the observed chemical composition of AGB. However, we note that, because of the rather large number of free parameters, AGB models are adjusted to match observations \citep{Lugaro16}. 
Therefore, a precise comparison between observed and predicted abundances is still mandatory to validate the different assumptions adopted in these complex models.

On the observational side, rather few studies are devoted to $s$-process element abundances in M-giant stars\footnote{We recall that the atmosphere of these M-type stars is still oxygen-rich, as it was when they formed.}, mostly because of the complex analysis of such cool star crowed spectra.
 \citet{Smith85, Smith86, Smith87, Smith88, Lambert91, Lambert95} analysed a few hundred stars in total and found a correlation between the $^{12}$C and $s$-process abundances at the AGB surface, in agreement with model predictions as both species are produced in the stellar interior. 
However, neutron sources in AGB, and especially in intermediate-mass (between $\sim$4 and $\sim$8\Ms), were still not fully understood. For low-mass AGB, the main neutron source is the $^{13}$C($\alpha$, n)$^{16}$O reaction. Nevertheless, by looking at the $s$-process element abundance distribution, some AGB stars show a flatter distribution compared to Solar one which led \citet{Danziger66} to suggest that much longer and/or larger density neutron exposures could occur in some AGBs \citep{IR83}. In fact, for intermediate-mass AGB, the main neutron source is now known to be the $^{22}$Ne($\alpha$,n)$^{25}$Mg reaction \citep{Cameron60, Iben75, Kappeler11}. Due to its higher neutron density, AGBs atmospheres can then be enhanced in Rb, $^{25}$Mg and $^{26}$Mg. However, some of these stars show no enhancement in Rb and no Tc signatures\footnote{Tc is a key element to prove the AGB nucleosynthesis as its isotope with the largest lifetime is $^{98}$Tc that rapidly decays after about 4.2 Myr.} are found which could be explained by the absence of a $^{13}$C pocket \citet{Lugaro16}. This is theoretically predicted by \citet{Goriely-Siess04, Herwig04} via the so-called Hot Dredge Up \citep{Straniero23}. 
In complement, some very specific studies concerning different sub-classes of AGB stars brought some complementary information about the AGB yields. One could, for instance, cite CH-stars \citep{Vanture92, Cristallo16}, Ba-stars \citep{Cseh22}, C-rich stars \citep{Utsumi70, Carlos02}, .... But the sample statistics were always rather small in all these studies. 
Therefore, there is still a lack of very large samples of AGB stars with homogeneous $s$-process abundances in order to better understand the yield productions of oxygen-rich AGB\footnote{Oxygen-richness being defined as C/O ratio smaller than unity}.


In this context, the spectroscopic observations collected by the ESA $Gaia$ mission are of particular interest. Indeed, thanks to the analysis of the \Gaia/Radial Velocity Spectrometer (RVS) spectra by the General Stellar Parametrizer from Spectroscopy module \citep[][\gspspec~hereafter]{GSPspecDR3}, chemo-physical parameters such as the effective temperature (\T), the surface gravity (\g), the global metallicity (\Meta) and the enrichment in  $\alpha$-elements with respect to iron (\AF), as well as up to 13 individual chemical abundances, have been determined for 5.6 million stars, including a few hundred thousand of AGB stars. Among those 13 chemical elements, three are formed via neutron capture processes: zirconium (Z = 40), cerium (Z = 58), and neodymium (Z = 60). Although the Zr line found in the RVS domain is not formed in cool star spectra, the Ce and Nd lines are well detected in the spectra of huge number of AGB stars, leading to the publication of a large catalogue of $s$-process abundances in these evolved stars.

The scope of this article is to focus on these Ce and Nd abundances in AGB stars analysed by the \gspspec~module. We recall that \gspspec~cerium abundances in less evolved stars have already been presented in \citet{GOAT23} to discuss the Galactic content and chemical evolution in this element. 
The present work is composed as follows. Section 2 presents the selection of the sample of AGB stars with high-quality Ce and Nd abundances, while Sect.~3 explores the general properties of this sample. We also included in this section a short overview of the Nd abundances in the Galactic Halo. We then discuss the observed content in Ce and Nd of AGBs by comparison with AGBs $s$-process model predictions. Finally, the conclusions of this work are summarised in Sect. 5. 

\section{The \gspspec\ sample of AGB stars with $s$-process element abundances}

Among the 5.6 million stars published within the \Gaia\ DR3 catalogue \citep{Vallenari22}, the \gspspec\ module \citep{GSPspecDR3} has been able to derive abundances of two second-peak $s$-process elements. In total, one can indeed retrieve 103,948 and 55,722 stars with Ce and Nd abundances, respectively, whatever the stellar type is. This first sample of 55,722 Nd abundances is referred as the {\it complete Nd sample}, hereafter. These numbers include abundances derived in AGB stars
but also in less evolved stars \citep[see, for instance for Ce,][]{GOAT23}. The scope of this section is to present the best working sample of these $s$-process element abundances for the best parameterised AGB stars.



\subsection{AGB stars parameterised by \gspspec}
\label{Sec:AGB sample}
We first remind that the \gspspec\ module has published atmospheric parameters and chemical abundances from the analysis of the $Gaia$/RVS spectra (R $\sim$ 11,500). The individual chemical abundances are estimated thanks to a specific grid of reference synthetic spectra and the GAUGUIN algorithm \citep{2012ada..confE...2B, RB16} considering Local Thermodynamical Equilibrium (LTE, hereafter), Solar abundances of \citet{Grevesse07} and the atomic and molecular
line data of \citet{BestArticleEver}. 

Together with the parameters and individual chemical abundances, \gspspec\ also provides quality flags (\flags, hereafter) which are recommended to select the most accurate parameters, including abundances. These flags can either be related to the stellar parametrization (induced, for instance, by possible biases caused by radial velocity uncertainties, flux noise or rotational velocities) or to the derived chemical abundance of a given element. The corresponding flags for the abundances of a $X$ species are $XUpLim$ and $XUncer$. We remind that the value "0" for all these flags corresponds to the best measurements and we refer to \citet{GSPspecDR3} for a more detailed description of these flags.

First, without considering any flag restriction and after applying the calibrations recommended by \cite{GSPspecDR3} for the atmospheric parameters\footnote{see~also~\url{https://www.cosmos.esa.int/web/gaia/dr3-gspspec-metallicity-logg-calibration}}, we found that, among all the \gspspec\ parameterised stars, 174,104 of them have a published  \T~$\le$~4000~K and \g~$\le$~2.0 (most stars cooler than $\sim$3650~K were disregarded for the DR3 analysis)\footnote{We adopt here the median value of the published parameters.}. These atmospheric parameter cuts were adopted in order to ensure the AGB nature of the selected stars, lately confirmed by their $K$-band absolute magnitude (see below).
Then, we considered only the best-parametrized stars by (i) setting all the first 13 flags equal to zero excepted the $extrapol$ flag that is fixed to be $\le$ 1 and (ii) selecting only stars having a \SNR >100 or a $gof$<-3.5\footnote{The $gof$ corresponds to the goodness-of-fit for the observed spectrum over the entire RVS spectral range.} (these last two criteria being linked to the detection of the Ce and Nd in the analysed spectra. See Tab C.9, C.10 and C.11 of \citet{GSPspecDR3}). 
The remaining sample of 128,335 cool-giant stars is called \AGB, hereafter. 

\begin{figure}
        \centering
        \includegraphics[scale = 0.25]{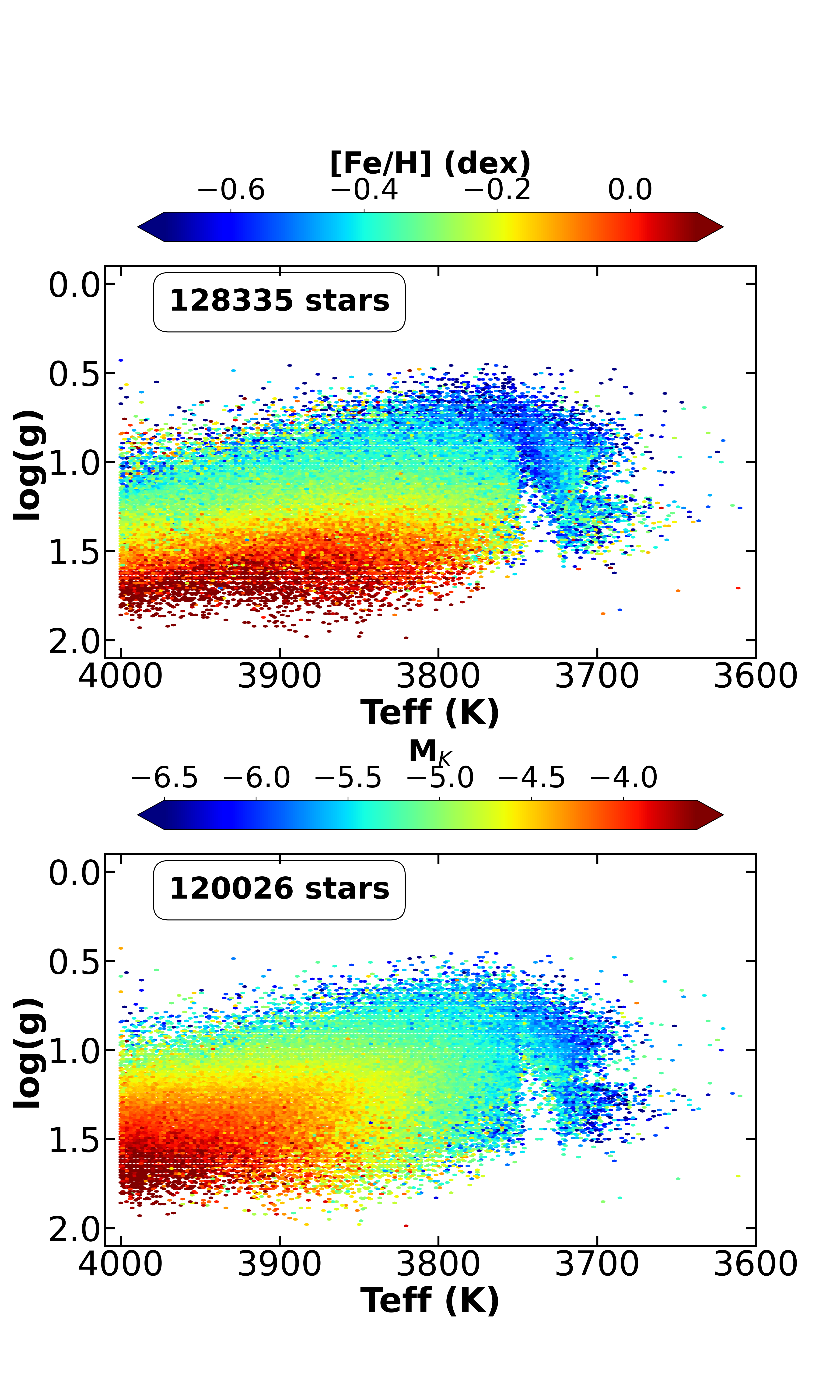} 
        \caption{Kiel diagram of the best-parametrized \gspspec~AGB stars colour-coded with their metallicity (top panel) and their absolute K magnitude (neglecting the extinction, bottom panel). This figure uses the calibrated \g\ and \meta\ as recommended by \cite{GSPspecDR3}.}
        \label{fig:AGB}
\end{figure}

The top panel of Fig.~\ref{fig:AGB} shows the Kiel diagram of these AGB stars colour-coded with their mean metallicity. It is noticeable that more metal-poor stars are found to have smaller gravities. We also remark a lack of stars around \T~$\sim$~3730~K. These biases result from the difficulty to parameterised more metal-rich and/or cooler AGB stars having complex spectra crowded by molecular lines. This has been explored with the $KMgiantPar$ flag (stars with a non-null value of this flag have set values of \T~and \g, see \citet{GSPspecDR3}). In order to reject such complex cases, we remind that we adopted for the selection criteria $KMgiantPar$=0.
 
In order to confirm the AGB nature of these \AGB~stars, we estimated their absolute $K$ magnitude value (M$_{K}$).
For that purpose, we adopted their apparent 2MASS $K$-band photometry \citet{Skrutskie06} and the photo-geometric distances from \citet{Coryn21}. We then computed the absolute $K$-magnitude for 120,032 stars (93.5\% of the whole sample) having a re-normalised unit weight error (RUWE) smaller than 1.4 and an astrometric fidelity factor for their astrometric solution ($fidelity\_v2$) larger than 0.5 \citep{Rybizki22}, assuring a good \Gaia\ astrometric solution. We note that, following this procedure, we neglect the interstellar extinction and thus, actually, derive a lower limit of the real M$_{K}$ (the stars are actually brighter than our estimate). We found that about 89\% of our stars are brighter than M$_K$ < -4.  This well confirms their AGB nature according to \citet{Abia22} who showed in their \Gaia -2MASS diagrams that AGB stars are typically brighter than M$_K \sim$ -4 to -5 mag (taking into account the ISM absorption and depending on their intrinsic \Gaia\ and 2MASS colours).  { Note that considering extinction only would make the stars a bit brighter. The resulting sample would probably consists in AGB stars in a larger fraction than the current one}. However, we can not completely exclude that our sample may be polluted by some extrinsic S-stars, CH-stars, R-stars\footnote{ R-stars spectra could be very similar to O-rich early-AGB or M stars} or early-AGB stars, that would have similar M$_K$ magnitudes \citep{Abia22}. {Such stars may not be enhanced in s-process elements and may introduce some dispersion in the observed [Ce/Fe] and/or [Nd/Fe] ratios with possibly negative values. }



The bottom panel of Fig. \ref{fig:AGB} shows the Kiel diagram of the \AGB\ colour-coded with their absolute K magnitude. We can see that cooler stars have lower gravities and larger M$_K$ confirming again their AGB nature. Finally, we note that 97\% of the working sample of AGB stars with Ce and Nd abundances (\AGBs~defined below) has M$_K$< - 4 mag (the other 3\% having M$_K$< - 3.2 mag), confirming their AGB nature.

\subsection{High-quality \gspspec\ Ce and Nd abundances in AGB stars}

Among the \AGB, \gspspec\ derived Ce and Nd abundances for 46,144 and 34,838 of them, respectively (without any abundance flags filtering). As the aim of this study is to analyse the largest sample with the most accurate chemical abundances, we present below a specific flag combination to build high-quality Ce and Nd working samples. 

\subsubsection{Cerium}
The description of the cerium line analysed by \gspspec\ and the associated sample of Ce abundances are already presented in \citet{GOAT23}. Briefly, high-quality Ce abundances were retained only if (i) Ce abundance uncertainties are smaller than 0.2~dex, (ii)  $vbroad \le$13~km/s, (iii) the abundance flags being set at $CeUpLim \le 2$ and $CeUncer \le 1$, (iv) $extrapol\le 1$ and, (v) $KMgiantPar \le 1$ if $gof$ < -3.75.
We remind that the $vbroad$ parameter provides information on the line broadening (rotational velocity, macroturbulence, ...) whereas $extrapol$ indicates if the \gspspec~atmospheric parameters are extrapolated beyond the reference grid limits. We refer to \citet{GOAT23, GSPspecDR3} for a complete description of the choice of these adopted filters. We also remind that it was found to be unnecessary to calibrate these Ce abundances \citep[see the discussion in][]{GOAT23}.
In the following, we only keep the \AGB~having the best Ce abundance as defined above, leading to a working sample of 17,765 AGB stars with high-quality cerium abundances (referred to \AGBCe, hereafter). Note that the Ce abundances of this sample vary from -0.17 to 1.10 dex and that no higher Ce abundances were found due to the cut in \T. We also recall that the $s$-process contribution for Ce is about 80\% at the Solar System formation epoch  \citep{Arlandini99, Bisterzo16, Prantzos18, Prantzos20}.

\subsubsection{Neodymium}
\label{Sec:Ne_Abund}

Neodymium abundances were estimated from one single line of \NdII~located at 859.389 nm in the vacuum (859.153 nm in the air)\footnote{The conversion between air and vacuum wavelength has been made using the \citet{1994Metro..31..315B} relation.}. Its lower level excitation energy and oscillator strength have been fixed to  1.357 eV and -1.650, respectively \citep{DenHartog2003}. We recall that neodymium has seven stable isotopes : $^{142}$Nd (27.13 \%), $^{143}$Nd (12.18 \%), $^{144}$Nd (23.80 \%), $^{145}$Nd (8.30 \%), $^{146}$Nd (17.19 \%), $^{148}$Nd (5.76 \%), and $^{150}$Nd (5.64 \%) \citep{DenHartog2003}. All these isotopes have been considered in the computation of the reference synthetic spectra. The $s$-process contribution for Nd is about 60\% \citep{Arlandini99, Bisterzo16, Prantzos18, Prantzos20}.
We finally note that this Nd line has not been astrophysically calibrated by \citet{BestArticleEver} because of lack of high-quality Nd abundances in reference stars. It could be slightly blended for some specific combinations of atmospheric parameters and chemical abundances by a weak CN line but, thanks to checks with synthetic spectra, this contribution has been found to be negligible for stars cooler than 4000~K (as long as the stellar atmosphere is not enhanced in carbon, see Sect. \ref{Sect:Carbon}). This Nd line could also be slightly blended by some weak TiO lines in rather metal-rich stars cooler than 4000~K. However, O and Ti being both considered as $\alpha$-elements, the TiO contribution is expected to be rather well modelled in the reference synthetic spectra as \alphaFe\ abundances are determined by the \gspspec\ pipeline before the estimation of the individual abundances (Ti and O abundances being assumed to vary in lockstep with \alphaFe).

\begin{figure}[t]
        \centering
        \includegraphics[scale = 0.15]{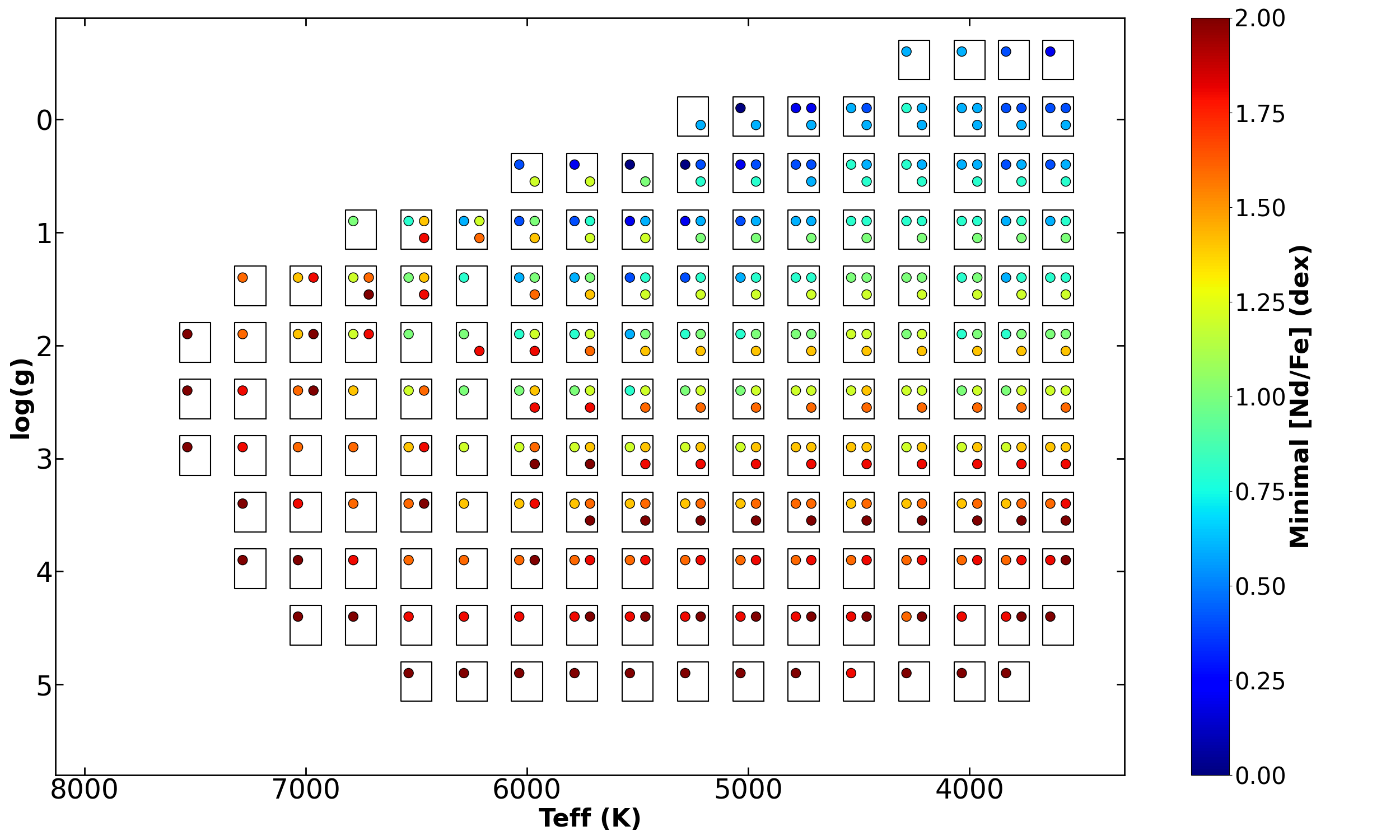} 
        \caption{Kiel diagram colour-coded with the lowest neodymium abundance (in dex) that could be detected in a spectrum whose (normalised) line centre flux is 0.5\% deeper than that of a reference spectrum computed with [Nd/Fe] = -2.00 dex. For each combination of effective temperature and surface gravity, each small square contains the estimate of the lowest Nd abundance for three values of [M/H] : 0.0, -0.5, and -1.0 dex (from top to bottom and left to right). }
        \label{fig:Nd-prof}
\end{figure}

First, as already presented by \citet{GOAT23} for cerium, we illustrate in Fig.~\ref{fig:Nd-prof}  the smallest Nd abundance (in dex) that could be measured for metallicities varying between -1.0 to 0.0~dex in a Kiel diagram. As expected, the Nd line is easier to detect in AGB stars (as far as \NdFe \ga 0.4~dex) than in dwarf stars (much higher Nd abundances are required to possibly detect this line).

Then, we build a subsample of AGB stars with the most accurate Nd abundances by applying the best combination of the \flags, following a similar procedure as for Ce. 
We first note that setting all \flags~=~0 (including those related to the neodymium abundances) results in a sample of five stars only. 
This number can be substantially increased while maintaining a high quality for the Nd abundances by  relaxing some of these \flags. 
For example, to build our working sample, we applied the same flag combination as for the \LUSCe\ (see Sect.~2.1.1). Note that we also kept stars with \NdFe$\ge$2.00~dex (abundance upper limit caused by border effects of the reference spectra grid) if their $KMgiantPar$ = 0 and $NdUpLim$ = 0. For these stars, we adopted in the following \NdFe=2.00~dex, although their abundances could be higher.
Our final working sample is then composed of 3,492 AGB stars with high-quality Nd abundance (which is referred to as \AGBNd, hereafter). An example of the ADQL query to retrieve this sample can be found in Appendix \ref{Append}.

As a validation check of the quality of these selected abundances, we look for literature Nd abundances for our \AGBNd\ stars. We found only a few stars in common with previous studies since none were exclusively devoted to AGB stars: six are in common with \citet{Taut21} (mainly FGK-type stars) and nine with \citet{BAWLAS}. Unfortunately, only nine of them have literature atmospheric parameters enough close to ours to allow a proper comparison of the Nd derived abundances (adopting differences in \T, \g\ and \meta \ smaller than 150K, 0.50 and 0.30~dex, respectively). Although our Nd abundances seemed to be systematically larger than the few found in the literature, {we decided not to calibrate them using a comparison sample due to the small number of stars}.  Nevertheless, as our sample is composed of giant stars only, we analysed the observed spectrum of Arcturus \citep{Hinkle00} with the \gspspec\ pipeline to estimate its Nd abundance, adopting the atmospheric parameters of \citet{BestArticleEver}. The rather larger \gspspec\ [Nd/Fe] in Arcturus being confirmed with respect to the value reported by
 \citet[][]{Fanelli21} (who adopted similar atmospheric parameters as us), we forced our [Nd/Fe] in Arcturus to be equal to theirs ([Nd/Fe]= - 0.03~dex in the \citet{Grevesse07} Solar scale) and calibrate all the other Nd abundances accordingly.

\begin{figure}[t]
        \centering
        \includegraphics[scale= 0.15]{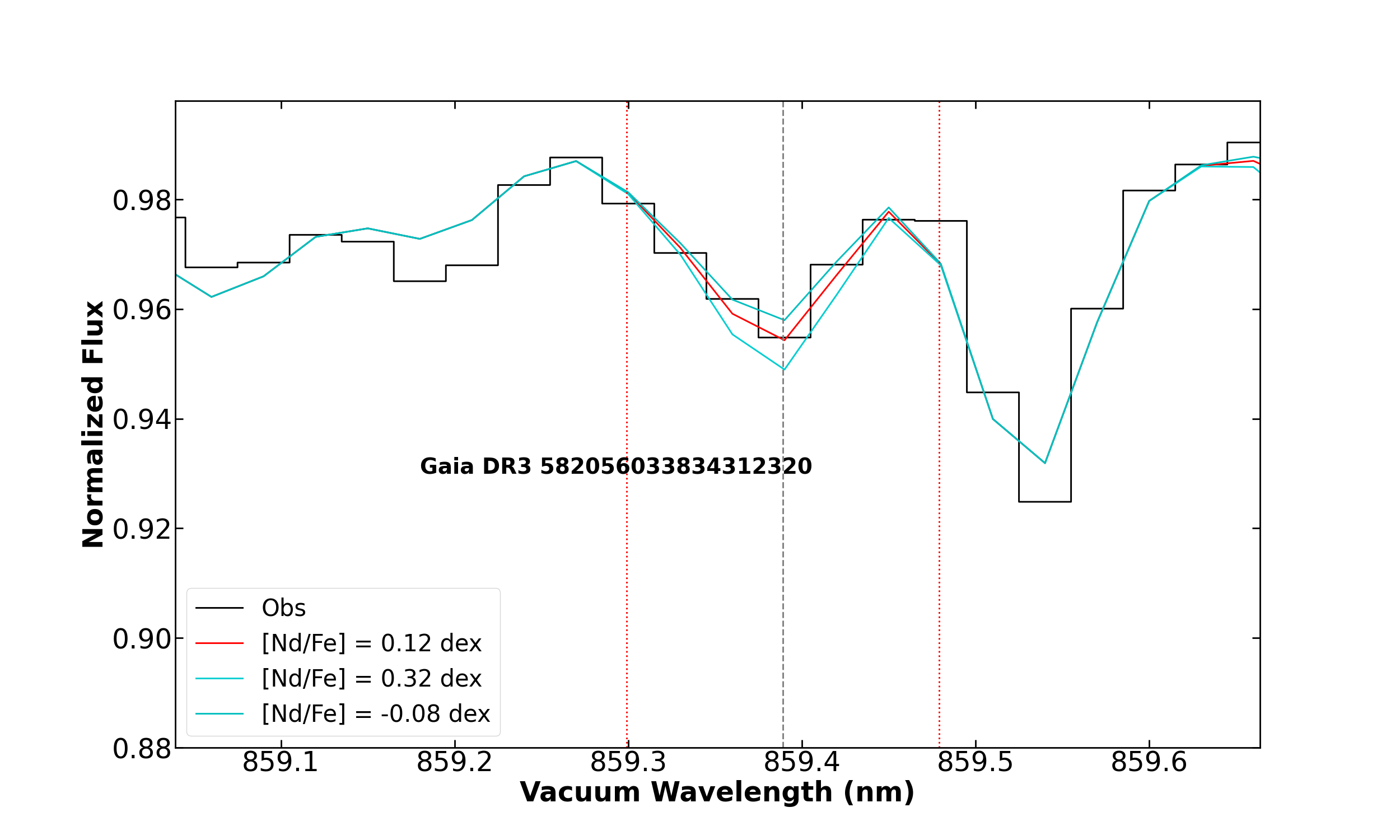} 
        \caption{Comparison around the \NdII~line between the $RVS$ public spectrum of $Gaia$ DR3 582056033834312320 (in black) and a model spectrum computed from its \gspspec\ parameters (in red). Cyan lines show synthetic spectra with Nd abundance $\pm$ 0.2 dex around the adopted Nd abundance. The calibrated parameters of this star are: \SNR~= 1110, \T~= 3825~K, \g~= 0.37, \FeH~= -0.74 dex, \AF~= 0.12 dex, \NdFe~= 0.12~dex and \CeFe~= 0.14~dex.}
        \label{fig:Nd-line}
\end{figure}

Finally, to illustrate the \gspspec\ analysis of these \AGBNd\ stars, we show in Fig.~\ref{fig:Nd-line} a comparison between the $Gaia$/RVS public spectrum and its associated synthetic spectrum, computed using the \gspspec\ atmospheric parameters and chemical abundances, for a star representative of the \AGBNd. As it can be seen, there is an excellent agreement between the observed (in black) and modelled spectra (in red), confirming the good determination of Nd-abundances in these AGB stars.

\section{Global properties of the $s$-process AGB sample}
\label{Sect:AGB}

We describe here the properties of the \AGBs\ which is composed by the 17,765 and 3,492 stars (the \AGBCe~and \AGBNd, respectively). As 1,713 of them have both Ce and Nd abundances, the \AGBs~is actually composed by 19,544 stars.


\subsection{Stellar types and oxygen-rich nature}
\label{Sect:Carbon}

We have already confirmed in Sect.~\ref{Sec:AGB sample} the AGB nature of the  19,544 stars that compose the \AGBs\ thanks to an estimation of their absolute magnitude in the $K$-band. 
Moreover, 445 of these stars have been classified as Long-Period Variables (LPV) by \cite{Leb22} thanks to Gaia/DR3 photometric epoch observations, leading to an independent confirmation of their AGB nature for a couple of them. Their mean period is $\sim$95~days and the longest periods reach $\sim$600~days (Gaia/DR3 data favouring the detection of LPV with the shortest periods).

Then, we have looked for the stellar type of these \AGBs\ stars.
In particular, we were interested in their possible atmospheric enrichment in carbon resulting from the dredge-up events characterising this specific stage of the evolution of low- and intermediate-mass stars. The presence of carbon (C-rich) stars in this sample could be problematic since their chemical analysis has been performed by assuming oxygen-rich stellar atmospheres and a Solar-scaled C/O ratio within the \gspspec\ module. This may slightly modify the stellar parameters and associated chemical abundances if different C/O ratios would have been adopted.
In any case, such a search of C-rich stars has been first conducted by checking the  $cnew\_gspspec$ parameter which is an indicator of a possible CN enrichment (or impoverishment) in the stellar atmosphere (C-rich star spectra being characterised by the presence of a strong CN line analysed by \gspspec).  We redirect the reader to \citet{GSPspecDR3} for a complete definition of this CN-parameter.
Among the 19,544 stars that compose our sample, 13,873 stars of them have a published $cnew\_gspspec$ value and all of them have this parameter close to 0, indicating that they have a close to Solar CN-abundance. Thus, they are not enriched in CN and this confirms their likely oxygen-rich nature. 
Moreover, we have also searched for possible carbon stars by looking at the C-rich flag proposed by \cite{Leb22}, leading to a list of about half a million of LPV candidates possibly enriched in carbon. Only 14 stars among the 19,544 of our sample could be suspected as being carbon-rich. However, this number should be considered with caution since this C-rich flag could lead to erroneous detection \citep[see][who reported about 1\% of false detections and see also the discussion in \cite{Sanders23}]{Messineo23}. On another hand, we also cross-matched our sample with the list compiled by \cite{Abia22} who examined the characteristics of already well-known Galactic carbon stars thanks to Gaia eDR3 data. These authors also presented a catalogue of about 2700 new Galactic carbon star candidates. None of our stars were found to be carbon-rich in this compilation but 18 of them were already known as being of spectral type $S$ (i.e. their C/O ratio being close to unity but they are still oxygen-rich). 

\begin{figure}[t]
        \centering
        \includegraphics[scale = 0.15]{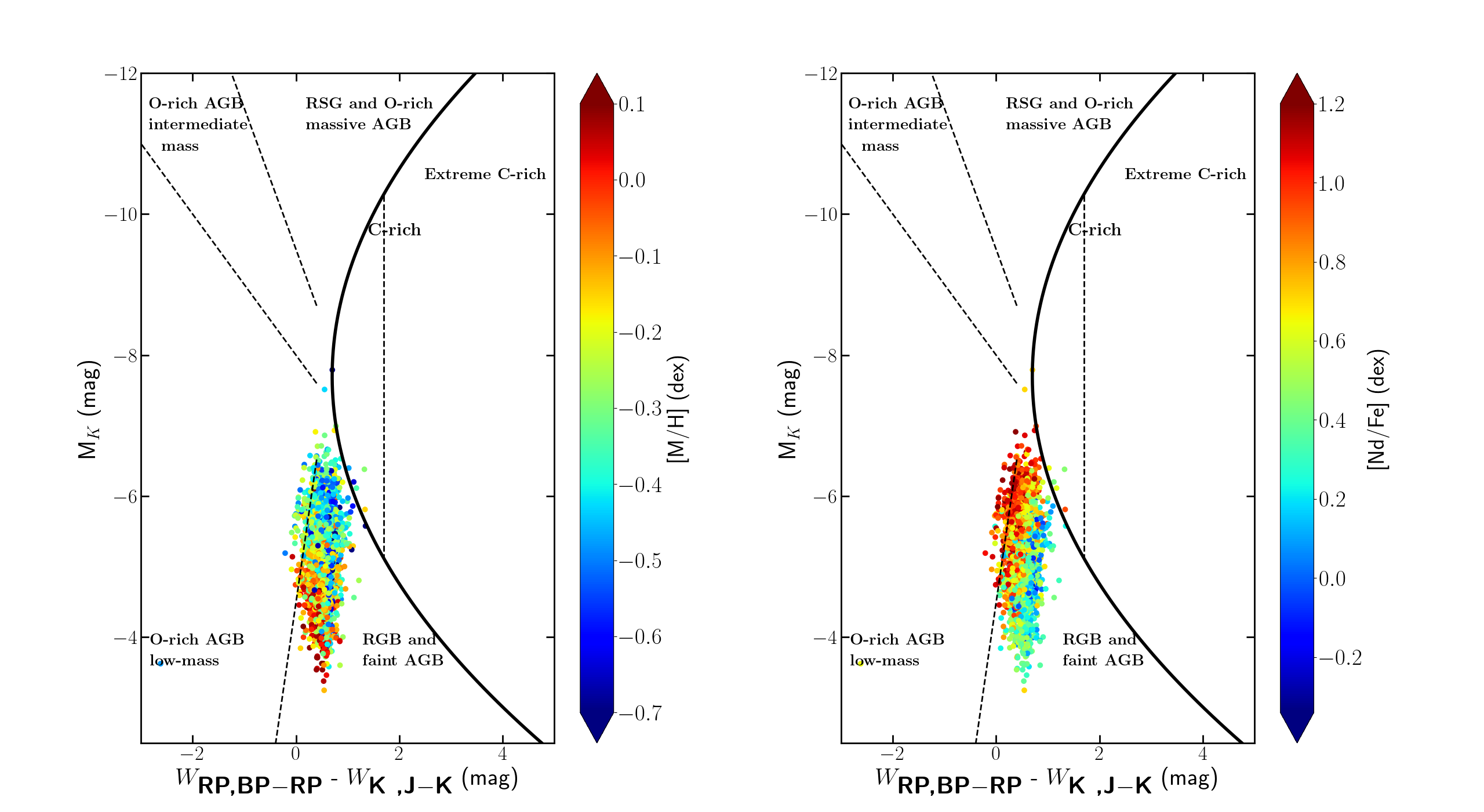} 
        \caption{\Gaia-2MASS diagram for the \AGBNd~colour-coded with metallicity (left panel) and Nd abundance (right panel). The curved line delineates the theoretical limit between O-rich (left-side) and C-rich AGB stars (right-side). The dashed lines separate sub-groups of stars as indicated in the figure.}
        \label{fig:Weh-Nd}
\end{figure}

Finally, to independently confirm the O-rich nature of our sample stars, we show the \AGBNd~in the \Gaia-2MASS diagram (Fig. \ref{fig:Weh-Nd}) colour-coded with the metallicity (left panel) and the Nd abundance (right panel). A very similar figure is obtained for the \AGBCe. This diagram shows the absolute $K$ magnitude with respect to a particular combination of the \Gaia\ and 2MASS photometry expressed through the reddening-free Wesenheit function \citep{So05} defined as W$_{\rm{RP,BP - RP}}$ = G$_{\rm{RP}}$ - 1.3(G$_{\rm{BP}}$ - G$_{\rm{RP}}$ ) and W$_{\rm{K ,J - K}}$ = K - 0.686(J - K) by \citet{Lebzelter18}.
We can clearly see that almost all the stars of this sample are located outside the C-rich regime, confirming their O-rich nature. Several of them also appears slightly fainter than typical AGB stars but this could be explained by the neglected extinction when computing their absolute magnitude. 
We finally remark that thanks to this \Gaia-2MASS diagram, we could have a hint of the stellar masses: they are found in the low-mass regime. This will be discussed later in Sect \ref{Sect:Discussion}.  

From all of these checks, we are thus confident about the real oxygen-rich and AGB nature of the \AGBs, strengthening the quality of their $s$-element abundance determination.


\subsection{Kinematical and dynamical properties} \label{Sect:Dyn}
These characteristics of the \AGBs\ were explored by using their Cartesian stellar coordinates, Galactocentric radii and cylindrical velocities, adopted from \citet{PVP_Ale} and the orbital parameters (such as Z$_{max}$ and eccentricity) from \citet{Pedro23}. \footnote{These parameters can be found in the table Performance verification/gaiadr3.chemical\_cartography in the Gaia archive \url{https://gea.esac.esa.int/archive/}). }

The large majority of the \AGBs~exhibits disc properties: 81\% of the stars have |Z$_{max}$|$<$1~kpc and about 88\% of our sample present eccentricities smaller than 0.25. The highest eccentricity values correspond to the lowest metallicities (\meta$<$-0.70 dex) hence the highest Ce and Nd abundances (see Fig. \ref{fig:Nd-Ce-Met} commented in the next subsection). Finally, about 88\% of the stars present a total velocity (quadratic sum of their three velocity components) smaller than 80 km/s confirming that the majority of the \AGBs~exhibits disc kinematical properties.  

\subsection{Chemical properties}
We present in this section the $s$-process abundances of these stars, first by focusing separately on the \AGBCe~and \AGBNd, and then, on the subsample of AGB stars having both Nd and Ce abundances.

\begin{figure*}
        \centering
        \includegraphics[width = 15 cm]{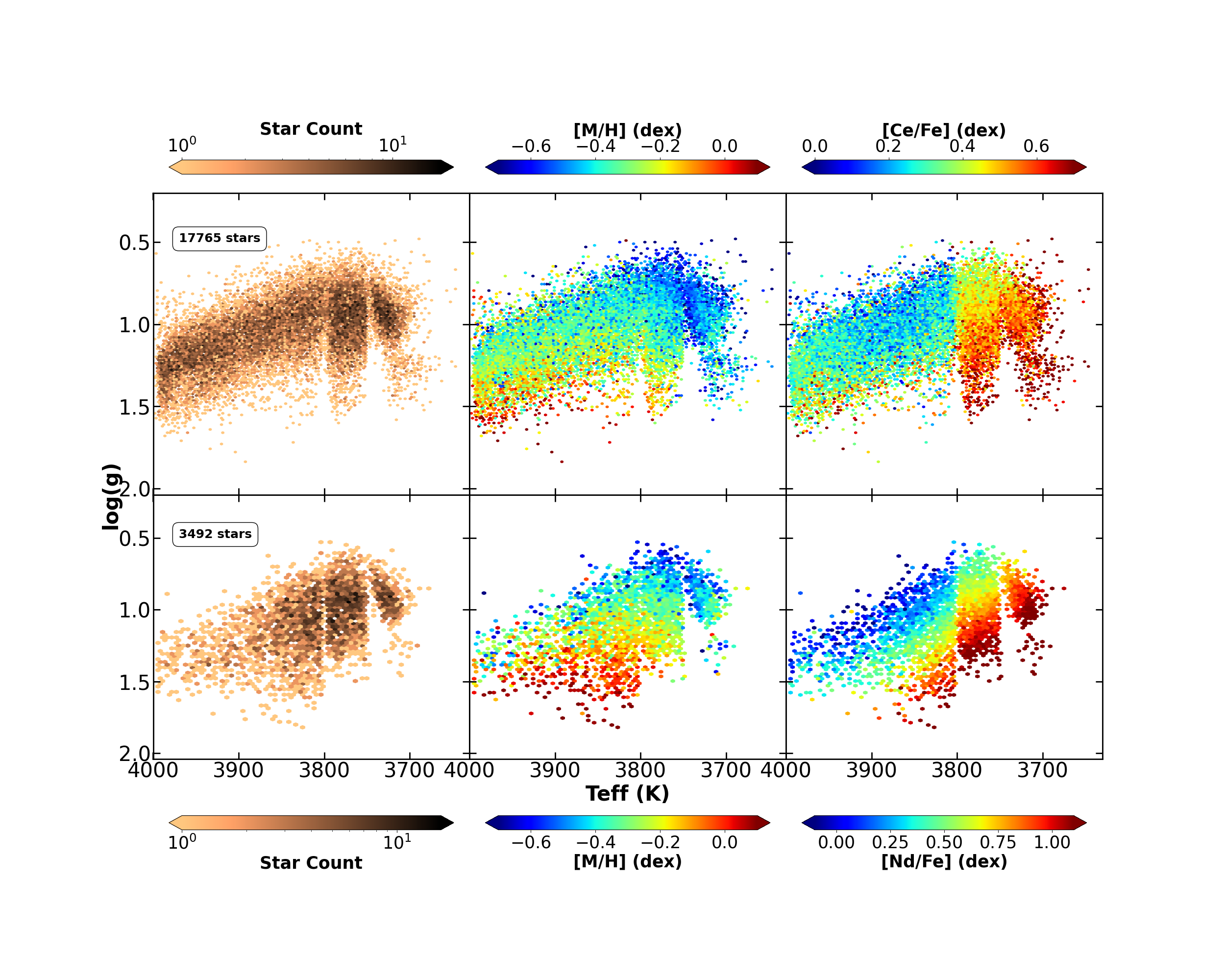} 
        \caption{Kiel diagrams of the \AGBCe\ (top panels) and the \AGBNd\ (bottom panels) colour-coded with the stellar count (left panels), the metallicity (central panels) and the corresponding s-process abundance (right panels). Note that this figure is an hexbin plot.}
        \label{fig:Nd-Ce-Kiel}
\end{figure*}

\begin{figure*}
        \centering
        \includegraphics[width = 15 cm]{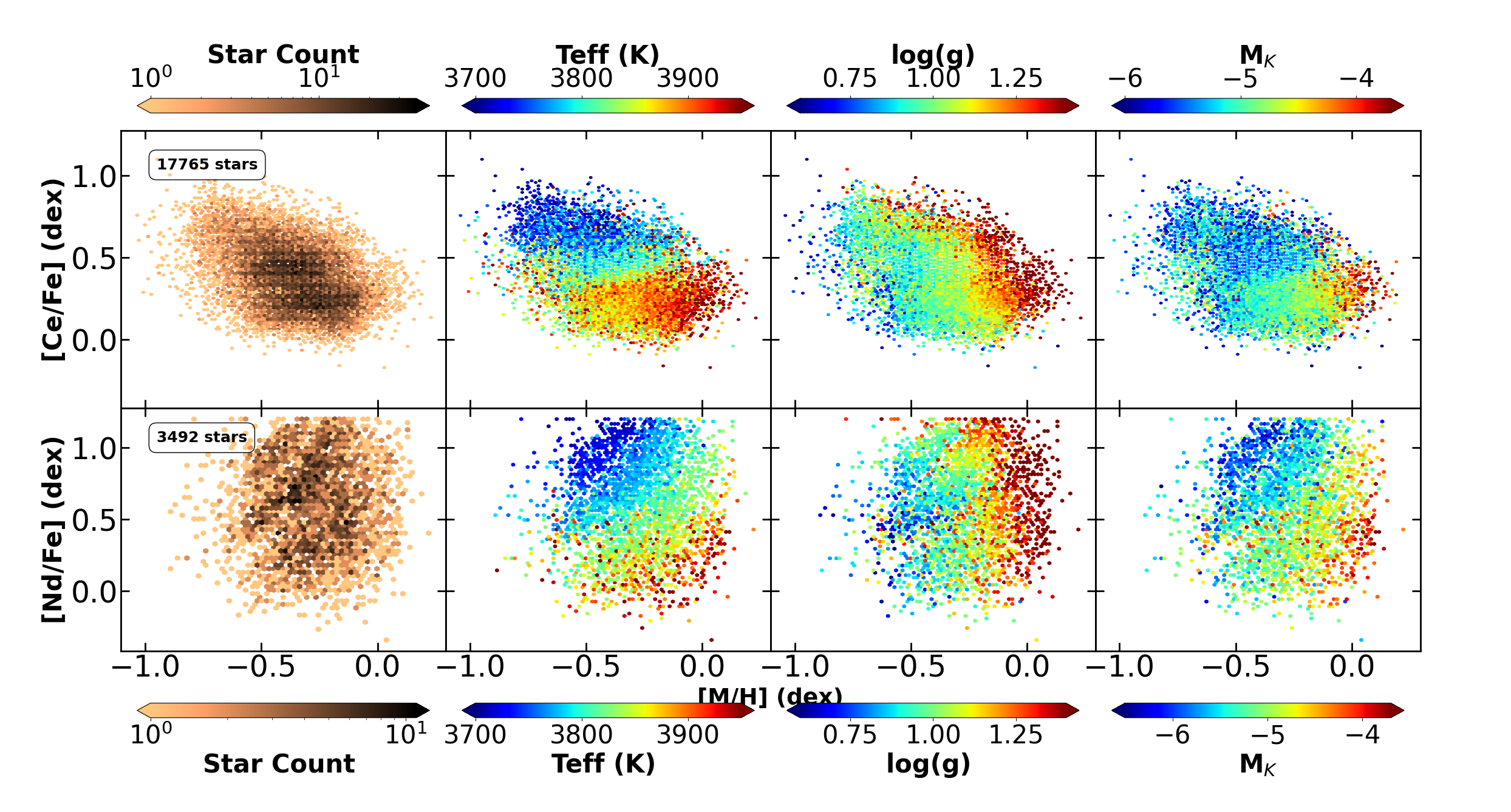} 
        \caption{Top panels : [Ce/Fe] with respect to metallicity colour-coded by (from left to right) stellar count, \T\, \g\ and M$_K$. Bottom panels: Same but for Nd abundances.}
        \label{fig:Nd-Ce-Met}
\end{figure*}

Fig.~\ref{fig:Nd-Ce-Kiel} first shows the Kiel diagrams for the \AGBCe~(upper panel) and \AGBNd~(lower panel) colour-coded with the stellar counts (left panels) and the metallicity (central panel). We remark that the metallicity of our sample stars decreases with decreasing \g. This observational bias is a general feature of the \gspspec~sample of AGBs as already discussed above (see Fig. \ref{fig:AGB} and associated text). The right panels show the Kiel diagrams colour-coded with Ce (upper panel) and Nd (lower panel). It is noticeable that at constant metallicity (hence a rather constant \g), the neodymium and cerium abundances are found to be larger for cooler and more luminous stars. Such features can also be observed by looking at the second and last column of Fig.~\ref{fig:Nd-Ce-Met} which presents the Ce (top panel) and the Nd abundances (lower panel) with respect to the metallicity colour-coded with \T~and M$_K$, respectively.  This can also be confirmed by looking at Fig.~\ref{fig:Weh-Nd}, where larger Nd abundances are found for more luminous stars. This is expected from AGB star evolution. Indeed, the successive TDUs characterising this evolutionary phase bring the material formed within the He-intershell towards the surface and, hence change the composition of the envelope. Initially enriched in O, the envelope becomes richer in s-process elements and primary carbon, which increases the radiative opacity and thus causes important changes in the physical structure of the AGB envelope. This impacts the stellar radius, luminosity, mass-loss, effective temperature as well as the formation of molecular species such as CN, HCN, C$_2$ \citep{Marigo02, Cristallo09}. At a given metallicity and \g, a cooler AGB has experienced more TP (hence more TDU) than when it was less evolved on the AGB and thus was slightly hotter, leading to an envelope more enriched in $s$-process elements as it can be seen in Fig.~\ref{fig:Nd-Ce-Kiel}. 

\begin{figure}[t]
        \centering
        \includegraphics[scale = 0.16]{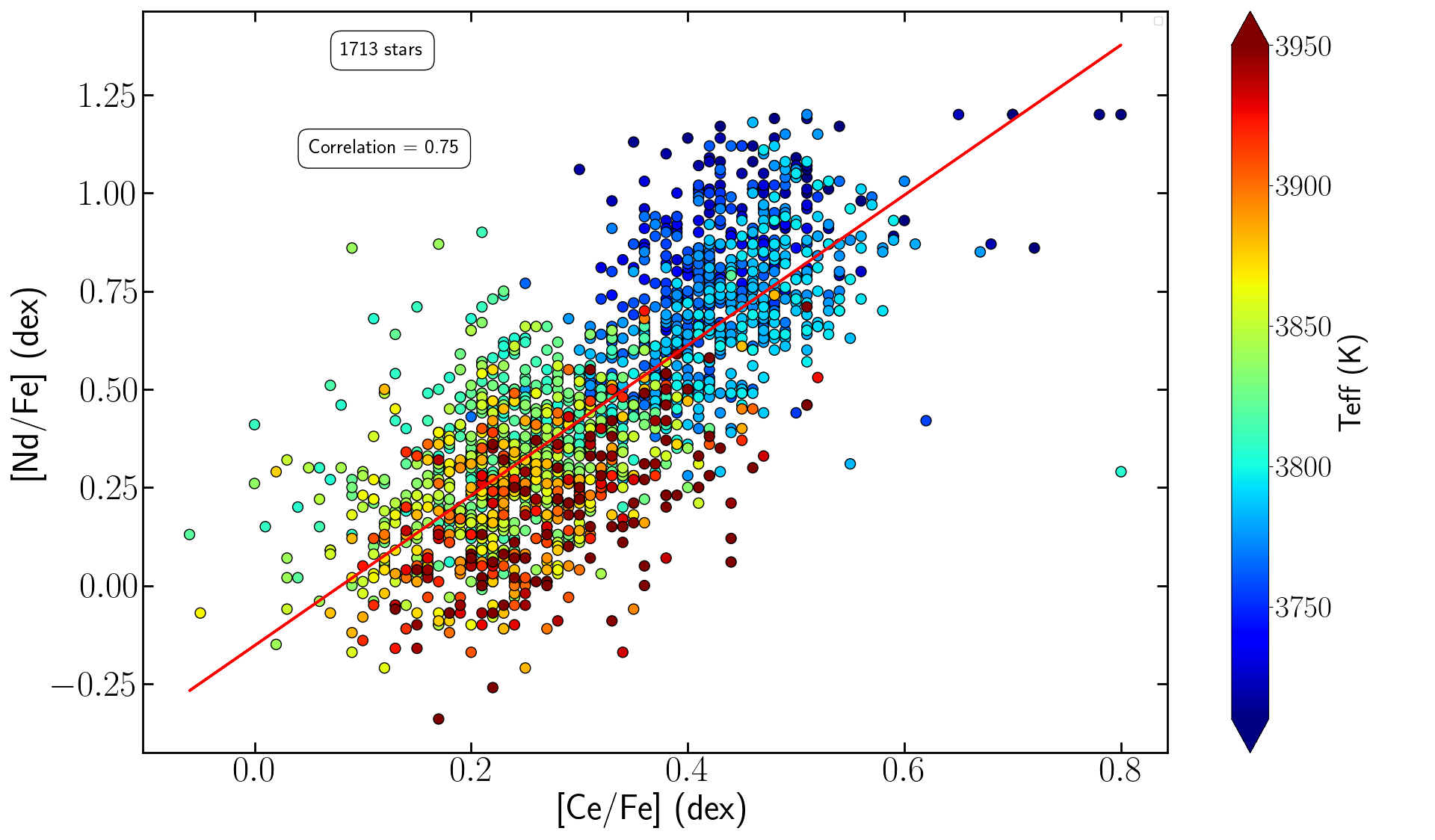} 
        \caption{Correlation between Ce and Nd abundances for AGB stars found in both the \AGBCe~and \AGBNd, colour-coded by their \T. The slope of the red line is 1.91 while the standard deviation is 0.44 dex. The Pearson correlation coefficient is also reported.}
        \label{fig:Corr}
\end{figure}

We also investigate the correlation between the \AGBCe~and \AGBNd~samples. We found 1,713 stars having both high-quality Nd and Ce abundances. As it can be seen in Fig.~\ref{fig:Corr}, a high positive correlation is found between those two chemical species (Pearson correlation coefficient  = 0.75)\footnote{ Note that there is a star having [Nd/Fe] = 0.23 dex and [Ce/Fe] = 0.80 dex. This outlier star is identified as S-Star.}. The associated dispersion is rather large since it is dominated by the Nd measurement uncertainties, the \NdII\ line being more difficult to analyse. Such a correlation confirms again the quality of the derived abundances since it is actually expected as both elements belong to the $s$-process second peak and hence have a similar nucleosynthetic origin. We also remark that colour-coding this correlation with \T\ clearly shows the enhancement of the Ce and Nd abundances for cooler stars, i.e. for more evolved stars on the AGB, as already illustrated in Figs.~\ref{fig:Nd-Ce-Kiel}~\&~\ref{fig:Nd-Ce-Met}.

\subsection{Neodymium in the Galactic halo}
Even though the majority of the \AGBNd~seems to belong to the disc, a small fraction of these stars belong to the Galactic Halo. Following the same procedure as in \citet{PVP_Ale}, we therefore investigate neodymium abundances in stars belonging to the halo component of the Milky Way (accreted systems and globular clusters).
We remind that a similar study for cerium in halo stars has already been presented in \citet{GOAT23}.

First, no stars in common were found by cross-matching the globular cluster compilation of the Harris catalogue \citep{Harris96} with the \AGBNd. Nevertheless, we found one star (ID=1821609090431019392) belonging to the M71 globular cluster when cross-matching with the lower-quality {\it complete Nd sample} (we recall that this sample corresponds to the 55,722 stars having Nd abundances but without any flag restriction). Note that we used a maximum separation in the sky of 0.5 degrees and check that the radial velocity of the stars is compatible with that of M71 with a difference lower than 0.5~km/s compared to the value estimated by \citet{Bau18}.
The (calibrated) metallicity of this star (\meta~= -0.72 $\pm$ 0.05 dex) is compatible with M71 literature estimates (-0.82 dex found in \citet{Carretta09} and -0.80 dex found in \citet{Boesgaard05}). The atmospheric parameters of this star are also very compatible with estimates of \citet{Sneden94, Greber20} with a mean \T~and \g~ differences of 70~K and 0.22~dex, respectively.
We found a Nd abundance of 0.95 $\pm$ 0.22 dex while no abundance of Ce has been found for this star. Such a value is higher than the mean [Ce/Fe] of M71 found in \citet{Masseron19} (0.27 dex) or the mean [Ba/Fe] (0.34 dex) and [La/Fe] (0.20 dex) found in \citet{Ramirez02}. This could be caused by the AGB nature of this specific star which may not be fully representative of the Nd abundance of this globular cluster and has an atmosphere already well enriched in $s$-elements.

We also explored the $Gaia$ DR3 Nd abundances of accreted systems, which have been identified as overdensities in the energy versus vertical angular momentum (E-L$_z$) diagram in \citet{PVP_Ale}. As no star was found to belong to any accreted system in the \AGBNd, we again investigate the {\it complete Nd sample} and reject all stars having a $KMgiantPar $ value larger than unity as well as all non-zero values for all parametrization flags.
We then found one star belonging to Thamnos \citep{Koppelman19, Helmi20}, four in the Helmi Stream \citep{Helmi20}, one in Sequoia \citep{Myeong19} and one in Gaia-Enceladus-Sausage \citep[GES; see][]{Helmi18, Belokurov18, Myeong18, Feuillet20, Feuillet21}. No chemo-physical literature studies have been found for those stars except for one in Helmi Stream (ID = 1591836174070974592) whose atmospheric \gspspec\ parameters are close to that of \citet{Jonsson20} (difference in \T, \g, \Meta, \AF = 31K, 0.32, 0.18 dex and 0.02 dex, respectively, while no Ce abundance was provided).
In order to investigate the accreted nature of these stars, we also kept only candidate stars with \Meta $<$ - 0.7 dex or having a low [Ca/Fe] ([Ca/Fe] $<$ 0.3 dex), similarly to the procedure adopted by \citet{GOAT23}. This selection removes three stars identified as Helmi Stream members. This misidentification is due to the proximity of the Helmi Stream to the solar neighbourhood in the (E-L$z$ diagram). Table \ref{Tab:Accreted} shows the $Gaia$ DR3 Id and the corresponding accreted system of the four stars as well as their atmospheric parameters, Nd and Ce abundances. 

First, the metallicity of the stars belonging to GES and Helmi systems are similar, as already found in \citet{GOAT23}. We remark a much lower metallicity for the star belonging to the Sequoia system compared to that of the three other accreted systems. A lower mean metallicity for Sequoia than for GES was already found in \citet{Feuillet21}. 

We also found a rather high [Ca/Fe] abundance for the star belonging to the Sequoia system. Even though smaller [Mg/Fe]\footnote{Mg and Ca being both $\alpha$-elements} abundances have been found in Sequoia than in GES by \citet{Matsuno22}, \citet{Feuillet21} found high-[Mg/Fe] stars belonging to Sequoia system. 

Moreover, we remark that the star belonging to GES presents a higher \NdFe~value than that of the Helmi stream (\NdFe = 0.83 dex), Sequoia system (\NdFe~= 0.35~dex) and Thamnos (\NdFe~=0.27 dex). As already suggested by \citet{RecioBlanco2021} with [Y/Fe] and \citet{GOAT23} with [Ce/Fe] measurements, a higher value of \NdFe~may suggest a higher mass of the system progenitor. This follows the mass estimations of these accreted systems by \citet{Koppelman19}. We note that \citet{Matsuno22} also found a higher abundance for the s-process elements (Y and Ba) in GES than in Sequoia. Finally, we note that the cerium abundance of the star from Thamnos system would suggest a similar mass of the progenitor to Helmi Stream. However, it is important to note that for the Thamnos and Helmi stream stars, the estimated \NdFe~and \CeFe~are compatible within error bars. However, our suggestions are based on only one star and should be taken carefully because of the adopted  Ce and Nd \flags. It is also important to note that this star of the Helmi stream may not be fully representative of the chemistry of the stream due to its AGB nature (being actually a producer of Nd). 


\begin{table*}[t]
        \centering
        \caption{\label{Tab:Accreted} \SNR, \T, \g, \meta, [Ca/Fe], \NdFe\ and \CeFe\ (and their associated uncertainties) for the four candidate accreted stars. Note that the \g~and \meta~and \CaFe~are calibrated according to \citet{GSPspecDR3}.}
        \begin{tabular}{lcccccccc}
                \hline
                 $Gaia$ DR3 Id & System & \SNR & \T & \g & \meta & [Ca/Fe] & \NdFe & \CeFe\\
                 & & & (K) & & (dex) & (dex) & (dex) & (dex) \\
                 \hline
                2410346779070638720 & GES & 68   & 4846 & 2.02 & -0.98 & 0.16 & 1.12 $\pm$ 0.24 & - \\
                1591836174070974592 & Helmi & 101 & 3806 & 0.95 & -0.73 & 0.37 & 0.73 $\pm$ 0.21 & 0.57 $\pm$ 0.20 \\
                5282079908816150912 & Sequoia &  407 & 4407 & 1.32 & -1.53 & 0.47 & 0.25 $\pm$ 0.19 & 0.36  $\pm$ 0.20 \\
                1294315577499064576 & Thamnos & 657   & 4309 & 1.51 & -1.01 & 0.27 & 0.17 $\pm$ 0.19 & 0.56 $\pm$ 0.08 \\

                \hline
        \end{tabular}
                
\end{table*}


\section{Constraints to AGB yield and evolutionary models} \label{Sect:Discussion}
The \AGBs\ presented above is an unique set of abundances that allows to constrain our understanding of the evolution of low- and intermediate-mass stars on the AGB. In particular, their yields for second-peak $s$-elements and their role on the Galactic chemical evolution of these species can be explored. 
In the following, we compare our abundances with yields predicted by FRUITY\footnote{available online via: \url{http://fruity.oa-teramo.inaf.it/}} \citep{Cristallo09, Cristallo11, Cristallo15} and Monash models \citep{Lugaro12, Fishlock14, Karakas16, Karakas18}. These two sets of models are the most complete in terms of explored masses and metallicities and allow an useful discussion on the impact of their different physical assumptions that could affect their yield prediction. 

\subsection{FRUITY and Monash models}

FRUITY (FUNS Repository of Updated Isotopic Tables and Yields) models are computed by considering simultaneously both stellar evolution and nucleosynthesis. In these models, the $^{13}$C pocket is formed in a self-consistent way and its mass is linked to an exponential decaying velocity function which includes a free overshoot parameter $\beta$. This parameter is set in order to maximise the $^{13}$C pocket and hence the $s$-process element production. Time-dependent convective overshoot at the base of the envelope is also included. In the following, we considered only non-rotating FRUITY models (as our stars have all $vbroad$ < 13 km/s).
We also adopted FRUITY models with a "standard" $^{13}$C pocket. 
In order to cover the metallicity range of the \AGBs, we considered five  metallicities (Z = 0.002, 0.003, 0.006, 0.010 and 0.014 corresponding to \meta~= -0.85, -0.67, -0.37, -0.15 and 0.00~dex, respectively) and eight  masses (1.30, 1.50, 2.00, 2.50, 3.00, 4.00, 5.00 and 6.00 M$_\odot$)\footnote{no lower masses being available}.

Monash model nucleosynthesis is computed thanks to a post-processing calculation based on the evolutionary models of \citet{Karakas14}. They include an exponentially declining profile of protons abundance in the top layers of the He-intershell
rather than directly including proton mixing into the He-intershell (as made in the FRUITY models). We recall that this proton ingestion has a capital impact on the formation of the $^{13}$C pocket (hence the $s$-process surface abundances) as these protons are captured by the present $^{12}$C to form $^{13}$C. Therefore, the size of the $^{13}$C pocket in Monash models is linked to a M$_{mix}$ parameter (defined as the mass of protons that are partially mixed over a mass extent in the He-intershell). For each Monash model mass and metallicity, we adopted the standard values of M$_{mix}$ provided in Tab.~2 and 3 of \citet{Karakas16}. Note that no rotation nor magnetic fields are included in these predictions. For the comparison with the observed abundances, we considered three metallicities (Z = 0.0028, 0.007, and 0.014 corresponding to \meta~= -0.70, -0.30 and 0.00~dex, respectively) and the same masses as for FRUITY, excepted that for Monash, the lowest considered masses were 1.0 and 1.25~M$_\odot$ (and not 1.3~M$_\odot$).

We note that in both sets of models, the mixing-length theory of convection is used and no convective overshoot is implemented before the AGB phase. The computations are stopped when the envelope reaches a critical mass but the adopted mass-loss relations differ. Finally, we note that these codes do not assume the same Solar abundances: FRUITY adopts the Solar scale of \citet{Lodders03} whereas Monash models are based on \citet{Asplund09} Solar abundances. In order to compare their predictions to our data, we scaled them to the Solar abundances of \gspspec~abundances \citep{Grevesse07}.

\subsection{Ce abundances as a proxy of the $s$-process content}
\label{Sect:Ce-Nd}

\begin{figure}[t]
        \centering
        \includegraphics[scale = 0.19]{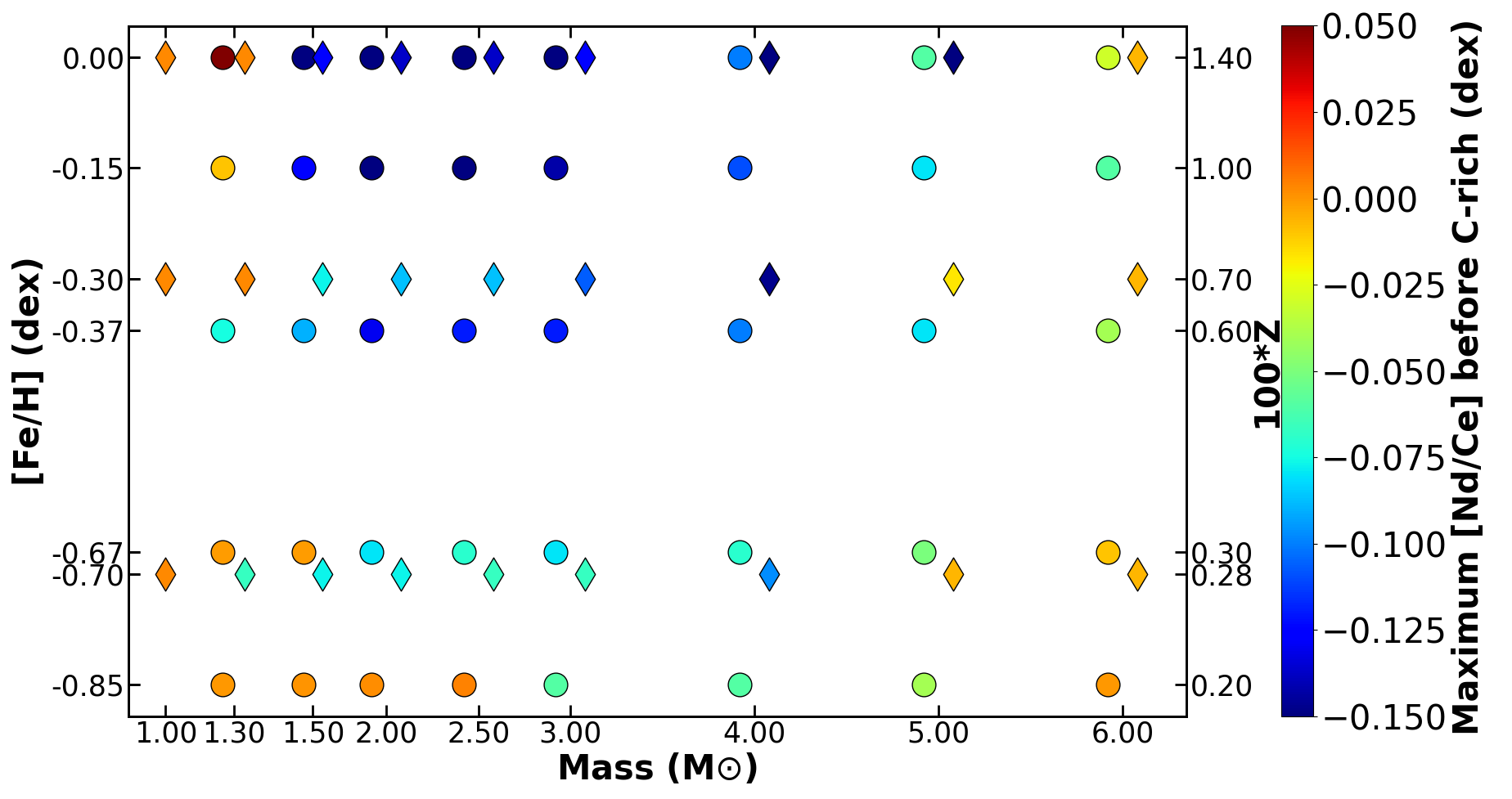} 
         \caption{[Nd/Ce] production predicted by the FRUITY (filled circles) and Monash models (filled diamonds) as a function of stellar mass and metallicity. The colour code corresponds to the maximum [Nd/Ce] production just before stars could become carbon-rich.}
        \label{fig:Comp-Models-NdCe}
\end{figure}

When stars start their AGB phase and before the first TDU episode, the Ce and Nd abundances were not affected by the previous dredge-up events. These species are thus still scaled to the Solar ones, i.e. [Nd/Ce]=0~dex. Then, the successive TDU enrich the AGB atmosphere in both elements. We have investigated the evolution of this [Nd/Ce] ratio along the AGB thanks to the  prediction of the FRUITY and Monash models, to check if the abundance of both elements are kept close each-other. For that purpose, we show in Fig.~\ref{fig:Comp-Models-NdCe} the [Nd/Ce] ratio predicted by both models at the surface of AGBs of different masses and metallicities before their atmosphere could become carbon-rich. This corresponds to the largest predicted Nd and Ce enrichments in an O-rich AGB. 
%
It can be seen that, for both sets of models, the absolute value of the maximum [Nd/Ce] ratio, before becoming a C-rich AGB, stays within 0.15 dex whatever the masses and metallicities are. This illustrates the rather constant [Nd/Ce] production rate in different types of AGB. This therefore confirms the correlation found observationally between both elements (see Fig. \ref{fig:Corr} and associated text). We thus decided to focus only on Ce hereafter in order to avoid our conclusions to be blurred by the larger uncertainties associated with our Nd abundances. Moreover, our results for Ce will be also valid for Nd.


\subsection{AGB model prediction of $s$-process elements  enrichment}
\begin{figure}[t]
        \centering
        \includegraphics[scale = 0.19]{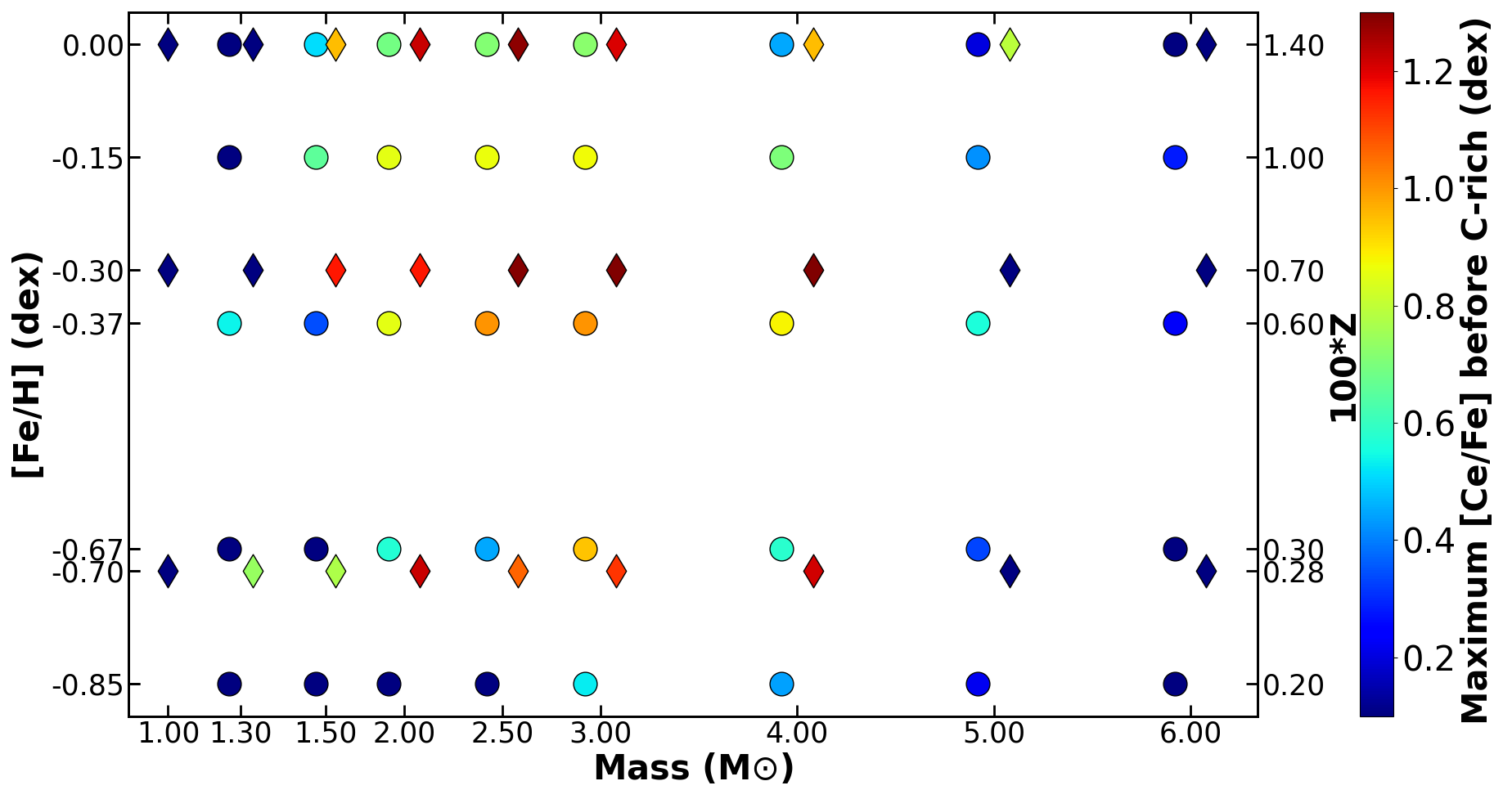} 
        \caption{Same as Fig. \ref{fig:Comp-Models-NdCe} but for [Ce/Fe] predicted abundances.}
        \label{fig:Comp-Models}
\end{figure} 

In order to explore the atmosphere enrichment in $s$-process elements along the AGB, we show in Fig.~\ref{fig:Comp-Models} the maximum \CeFe~abundance ratio of an oxygen-rich AGB in a similar way as in Fig.~\ref{fig:Comp-Models-NdCe}.
It can be remarked that for both sets of models, AGB stars of initial masses between 2 and 4 \Ms\ and metallicities  between -0.80 dex and -0.10 dex present the largest Ce enrichment at their surface, exceeding [Ce/Fe]$\sim$ 0.70~dex. 

Briefly, we recall that, for such masses (but this also applies to lower ones), second-peak $s$-process elements (such as Ce and Nd) are produced in the AGBs interiors, within the so-called He-intershell (mainly composed of He and $^{12}$C). Successive penetrations of the convective H-rich envelope (the so-called TDU) in this layer bring the material formed in the He-intershell towards the surface, changing drastically the envelope chemical composition.
During the interpulse, protons left by the receding convective envelope are captured by the $^{12}$C present in the He-intershell. This produces $^{13}$C through the $^{12}$C(p, $\gamma$)$^{13}$N($\beta^+ \nu$)$^{13}$C reaction and forms the so-called $^{13}$C pocket.  This $^{13}$C is then converted into $^{16}$O through the $^{13}$C($\alpha$,n)$^{16}$O reaction and provides an important amount of neutrons, which finally activated the $s$-process production \citep{Straniero95}\footnote{Another source of $^{13}$C is present in the top layers of the He-intershell because it is produced in the H-burning shell. However, this source has a negligible impact on the $s$-process production since most produced neutrons are captured by the abundant $^{14}$N element.}. 
Note that $^{14}$N and $^{22}$Ne pockets are also formed and partially overlap the $^{13}$C pocket. As $^{14}$N is a neutron poison because of the $^{14}$N(n,p)$^{14}$C reaction, the $s$-process nucleosynthesis occurs in the more internal tail of the $^{13}$C pocket where the amount of $^{14}$N is low \citep{Cristallo09}. 
All these elements formed in the He-intershell are brought up in the convective envelope by the subsequent TPs and TDUs. Moreover, the TDU has its maximum efficiency for this mass range between 2 and 4 \Ms\ \citep{Cristallo15}. As already noticed by \citet{Cristallo11}, we also highlight that, since intermediate-mass AGB stars have the largest production of Ce, they should play a large contribution to the interstellar medium enrichment, especially those with a mass around 2~\Ms\ because they are also more numerous  and thus provide a good balance between Initial Mass Function of the Galaxy and AGB Ce production rate \citep{Karakas16}.

We also note that, although their global conclusions about Ce production are rather close, it can be seen that Monash models (at any mass and metallicity) predict slightly larger Ce than FRUITY. For instance, for all masses below 4 \Ms\ and a metallicity around -0.35~dex, Monash models predict a Ce production 0.32 dex higher than FRUITY. This is caused by a larger number of predicted TP in Monash models than in FRUITY, resulting from the different adopted mass-loss relations \citep{Karakas16}.

On the other hand, we also remark in this Fig.~\ref{fig:Comp-Models} that, for the two largest considered masses (5 and 6 M$_\odot$), the Ce production predicted by both sets of models is much lower than for less massive AGBs (at all metallicities). In such massive AGB stars, the $^{22}$Ne($\alpha$,n)$^{25}$Mg neutron source indeed starts to be activated as higher temperatures are reached at the base of the convective zone of the AGB. As indicated in \citet{Cristallo15}, the neutron exposure of $^{22}$Ne($\alpha$,n)$^{25}$Mg is smaller than that of the $^{13}$C($\alpha$, n)$^{16}$O source. There is thus a considerable reduction of the second peak $s$-process elements production.
A decrease of the TDU efficiency furthermore accentuates this phenomenon. In fact, as the mass of the exhausted H-core is larger, this implies a larger compression of the H-exhausted layers and a thinner and hotter He-intershell. Therefore, a shorter time is needed to reach the ignition condition for activating the 3$\alpha$ reaction, leading to a shorter interpulse episode and hence a less efficient TDU \citep{Straniero03, Cristallo15}. 
However, it is interesting to note that almost no Ce is produced in the higher-mass AGBs of Monash models whereas FRUITY models predict slightly higher abundances by about 0.15~dex for initial masses between 5 and 6 \Ms. 
This is due to the fact that no $^{13}$C pocket is included in the Monash models for such high masses, whereas it is included for smaller masses. Hence, the source of neutrons in these models is only $^{22}$Ne($\alpha$,n)$^{25}$Mg which preferentially creates first peak $s$-process elements. Such a reaction is also more efficient in the latest TP \citep[while the $^{13}$C($\alpha$, n)$^{16}$O reaction is more active in the first TPs, see][]{Karakas16}.
All of this explains the absence of Ce production in Monash models for such high masses.



Regarding now the dependency of the Ce production with metallicity, we remark that AGBs with the lowest metallicity of the FRUITY models considered in this work (Z = 0.002 corresponding to \FeH = -0.85~dex) enrich the atmosphere up to \CeFe $<$ 0.30 dex (before the atmosphere becoming C-rich) only, except for the masses between 3 and 4~\Ms\ for which \CeFe~reaches 0.50 dex. At such low metallicities (\FeH~< - 0.80 dex), there are indeed less heavy seed nuclei available for captured neutrons, which limits the production of heavy elements like cerium. Note that, for even lower metallicities, the production of Ce and Nd is predicted to be larger since the number of available neutrons per seed nuclei (Fe) becomes also larger. Such conditions also favour the production of Pb and Bi.


 

\subsection{Comparing observed and model Ce abundances}

We now compare the Ce abundances derived for the \AGBCe~with the theoretical predictions of FRUITY and Monash models of 5 and 6, 2.50 and 1.50 \Ms, in light with the previous subsections describing both models.  We recall that similar results can be found using the \AGBNd.

We first begin this comparison for stars having the two highest masses.
As already illustrated and discussed in Fig.~\ref{fig:Comp-Models}, the most massive AGB stars in both models do not produce [Ce/Fe] abundance ratios larger than $\sim$0.3 dex. Therefore, the larger observed abundances have to be explained by invoking lower stellar masses. This independtly confirms the probable absence of massive AGB stars in our sample as already noticed in the \Gaia-2MASS diagram shown in Fig. \ref{fig:Weh-Nd}.

\begin{figure}
        \centering
        \includegraphics[scale = 0.17]{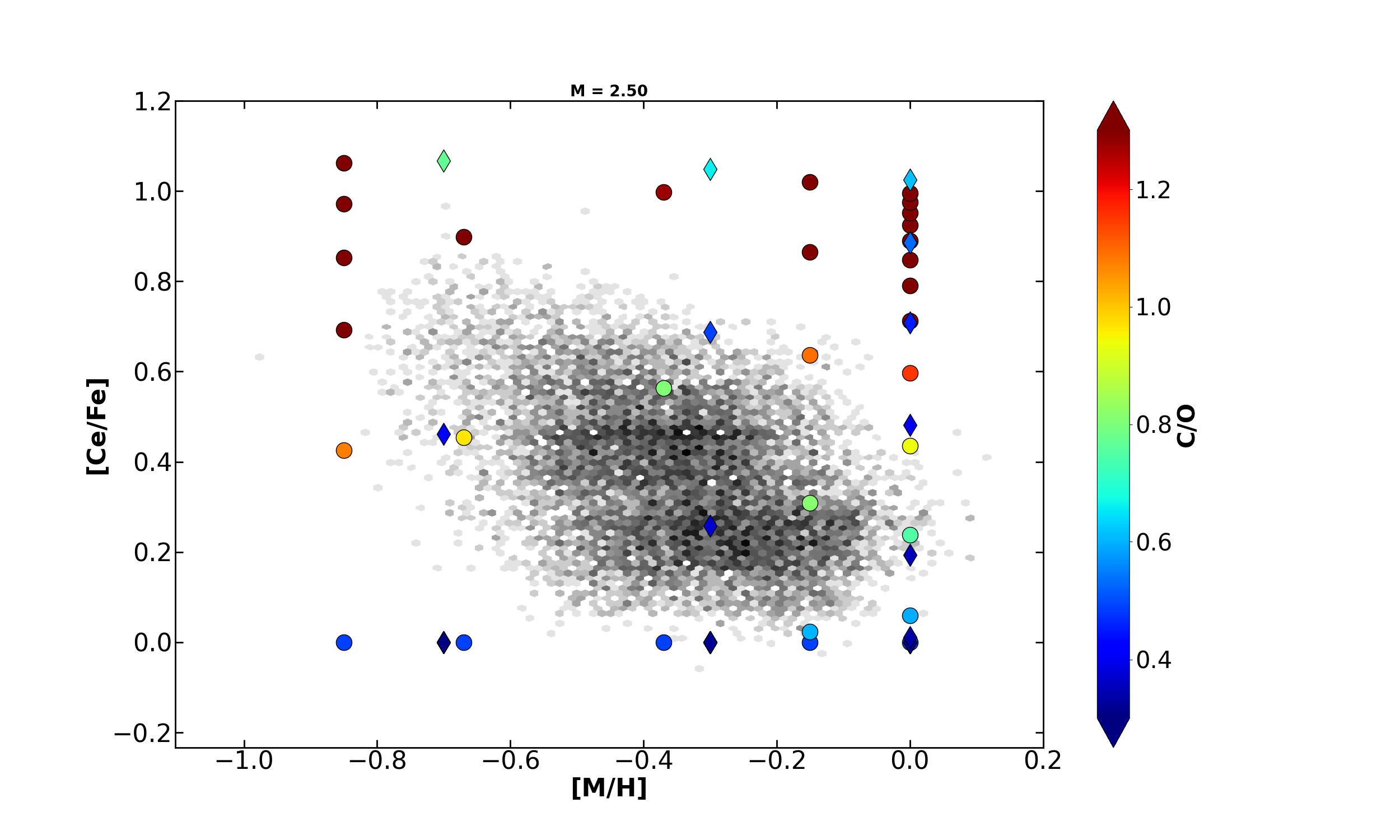} 
        \caption{\AGBCe~colour-coded by stellar counts. Filled diamonds (Monash models) and circles (FRUITY ones) correspond to the [Ce/Fe] predicted after each TDU, colour-coded by the C/O ratio. The AGB mass considered here is 2.50 \Ms.} 
        \label{fig:Mod-vs-Obs}
\end{figure}

We then focus on the AGB stars with an initial mass of 2.50 \Ms\ since similar conclusions can be obtained for stars of initial mass between 2 and 4 \Ms. 
Fig.~\ref{fig:Mod-vs-Obs} shows 
for FRUITY and Monash model the \CeFe\ ratio predicted after each TDU, colour-coded by the C/O abundance ratio.
We first remark that almost no \AGBCe\ stars are found where models predict the formation of carbon-rich stars, in full agreement with our conclusions about the oxygen-rich nature of our sample stars.
Then, it can be seen that the [Ce/Fe] ratio predicted by both models are fully compatible with the observed values. Therefore, for metallicities between -0.80 and 0.00~dex, the yields predicted for AGBs with an initial mass of 2.50~\Ms~by both models are fully compatible with the observed \gspspec~Ce abundances that are found between 0.00 and 0.80 dex. 



\begin{figure}
        \centering
        \includegraphics[scale = 0.17]{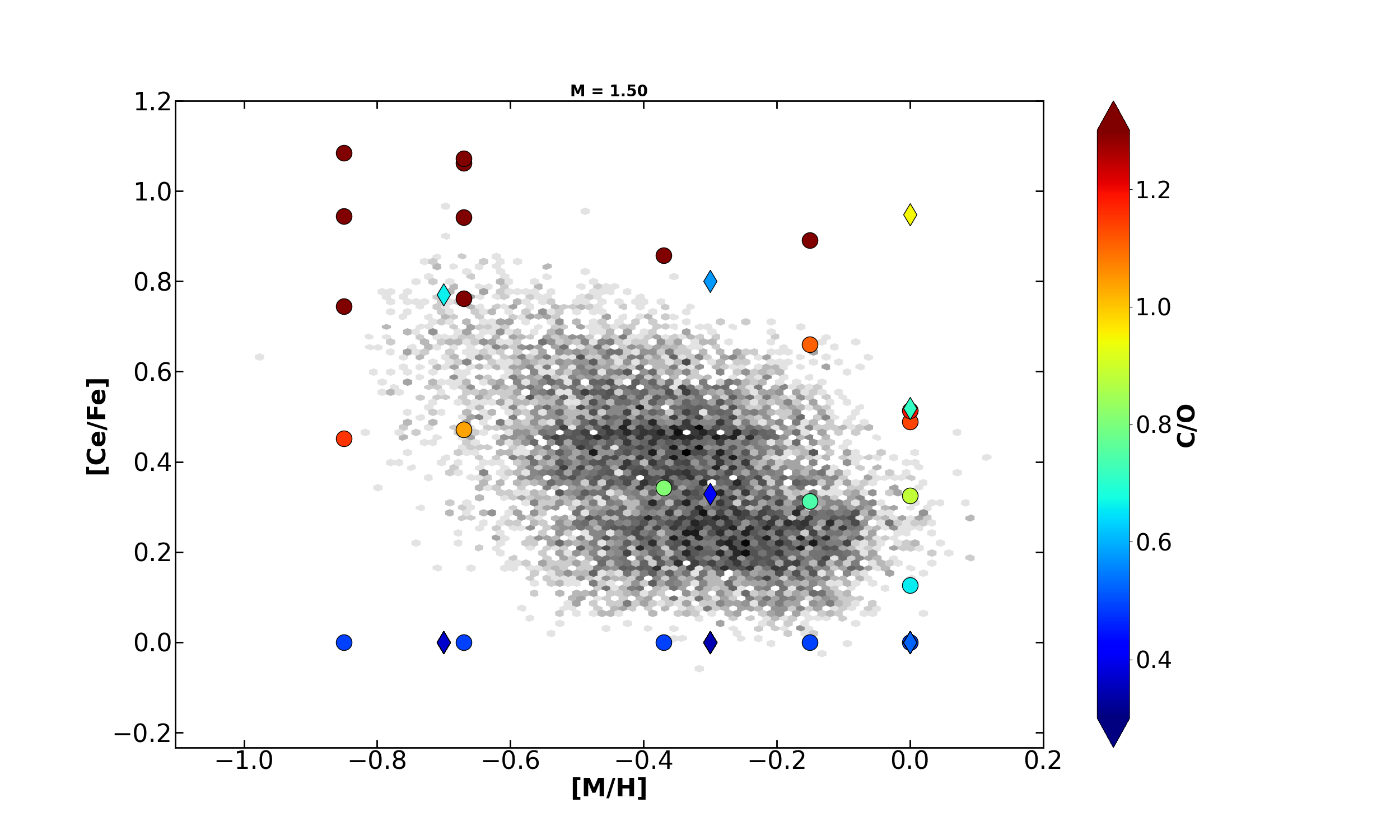} 
        \caption{Same as Fig. \ref{fig:Mod-vs-Obs} but for a mass of 1.50 \Ms.}
        \label{fig:Mod-vs-Obs-M15}
\end{figure}

Fig.~\ref{fig:Mod-vs-Obs-M15} is similar to Fig.~\ref{fig:Mod-vs-Obs} but for a mass of 1.50 \Ms. For metallicities between -0.40 and -0.10~dex, none of the stars of the \AGBCe~is located where models predict that the AGBs atmosphere is C-rich. 
However, for metallicities around -0.70 dex, we remark that Monash models predict that the AGBs are still O-rich whereas FRUITY models predict a C-rich atmosphere. As our stars are not C-rich, it seems that FRUITY models predict a too-large C abundance at the surface of the AGBs of this mass and metallicity. A similar conclusion can be drawn by looking at stars around Solar metallicity.
However, we note that we do not detect any stars with Ce abundance between 0.60 and 0.80~dex at Solar metallicity whereas Monash models predict such high values for  O-rich AGBs. This could be explained by (i) either the difficulty of automatically detecting the Ce line in RVS crowded spectra of cool metal-rich stars and/or (ii) some Ce overproduction predicted by Monash that are not confirmed by FRUITY.
Moreover, the spread of our \CeFe~vs~\Meta~could be explained because the AGB stars of our sample have experienced a different number of TDUs. They may also have a large variety of masses and different values of \AF~for a given metallicity. This is also due to the intrinsic dispersion in the formation of the $^{13}$C pocket (different profile and masses), even for samilar mass and metallicity stars. Finally, such a spread could also be partially explained by some stars which can be Ba-stars or CH-stars, as mentioned in Sect. \ref{Sec:AGB sample}, and thus, the Ce and Nd dispersions could be the result of the dilution effect of the accreted material in the secondary star.

However, heavy $s$-process elements such as Ce and Nd abundances alone are not enough to disentangle this large variety of stellar parameters.
For instance, AGB stars of lower metallicities (\FeH $<$ -1 dex) and masses (less than $\sim$3\Ms) produce large amount of Pb ([Pb/Fe] $>$ 2 dex) \citep{Lugaro12, Straniero14, Fishlock14, Cristallo15}.
Other constraints could also be obtained by looking at light $s$-elements such as Sr, Y and Zr. For instance, the [Ce/Y] ratio should decrease with increasing [Fe/H] due to the neutron exposure in the $^{13}$C pocket is proportional to the $^{13}$C/$^{56}$Fe ratio. The number of iron seeds is scaled to the metallicity whereas the $^{13}$C is not because of its primary origin \citep{Busso01}. On the other hand, the production of light $s$-elements should increase with increasing mass because of the activation of the $^{22}$Ne($\alpha$,n)$^{25}$Mg neutron source.
Indications of the main neutron source (hence the mass) could also be obtained by looking at the [Rb/Zr] ratio \citet{Abia01, Karakas16, Shejeelammal20, Roriz21}. If this ratio is negative, the main source is $^{13}$C($\alpha$, n)$^{16}$O which provides a relatively low neutron density. As already discussed above, this source is mainly active in low-mass AGB stars. On the other hand, a positive [Rb/Zr] ratio is caused by the $^{22}$Ne($\alpha$,n)$^{25}$Mg which produces a higher neutron density which opens branching points and produces Rb.

Therefore, further analysis of complementary $s$-process element abundances in this \AGBs~ may be useful to better understand and constrain the nucleosynthesis and atmospheric enrichment predicted by AGBs models.



\section{Summary}
The production of $s$-process elements in AGB stars parameterised by the \gspspec~module from their Gaia/RVS spectra has been studied. We first confirmed the AGB nature of these stars by looking at their absolute magnitude in the $K$-band using the photo-geometric Gaia distances \citep{Coryn21} and 2MASS photometry, and also their location in a \Gaia-2MASS diagram. Their oxygen-rich nature has also been confirmed. We then defined a high-quality sample (\AGBs) of 19,544 AGB stars with Ce and/or Nd abundances after applying a specific \gspspec~flag combination. 
The Nd abundances were calibrated thanks to the analysis of the Arcturus spectrum while no calibration were applied for Ce abundances according to \citet{GOAT23}.


We then investigated the kinematical and dynamical properties of this \AGBs~and found that the majority of the stars seems to be located in the Galactic disc. They have indeed a rather low total velocity ($<$ 80 km/s) and a Z$_{max}$ lower than 1.0~kpc. We also found one Halo star belonging to the M71 globular cluster. Its atmospheric parameters are fully compatible with literature studies but a higher Nd abundance than M71 mean $s$-elements abundances is found. This could be explained by the AGB nature of this star, enriched in Nd by successive Dredge Up events. We also found four stars belonging to the accreted systems Helmi, Sequoia, GES and Thamnos. The GES star is found to have a higher Nd abundance than those of the other systems. This higher Nd abundance could be linked to a higher mass of the progenitor. Such a correlation between $s$-process abundances and progenitor mass was already highlighted with Y in \citet{RecioBlanco2021} and Ce in \citet{GOAT23}.

The chemical properties of our sample revealed that, at a given metallicity, the Ce and Nd abundances are higher for cooler stars. This is an expected feature of the AGB stars and shows the progressive enrichment of their atmosphere in $s$-process elements when more TP and mixing events occur. Moreover, 
among the 19,544 stars of the \AGBs, 1,713 stars have both high-quality Ce and Nd abundances, that are very well correlated. This is expected as those two elements belong to the same $s$-process peak and hence have a similar nucleosynthetic origin, showing again the high-quality of the \gspspec~abundances.

The comparison between the \gspspec\ $s$-process abundances and predicted AGBs yields by the FRUITY and Monash models was then presented. These models allowed first to confirm the good correlation found observationally between Ce and Nd abundances. 
We then examined both models and found that AGB stars of initial mass between $\sim$2 and $\sim$4~\Ms\ and metallicities between $\sim$-0.80 and $\sim$-0.10 dex are the main Ce producers, reaching [Ce/Fe]~>~0.70~dex. In such stars and in those of lower mass, the main neutron source is $^{13}$C($\alpha$, n)$^{16}$O and preferentially forms second peak $s$-process elements such as Ce and Nd. However, Monash models predict slightly higher Ce abundances compared to FRUITY ones due to different physical and computational assumptions.
On the other hand, the Ce production in more massive AGBs ($\sim$5-6 \Ms) is much lower as the main neutron source is $^{22}$Ne($\alpha$,n)$^{25}$Mg, leading preferentially to the formation of first-peak $s$-process elements.

Then, we confronted these models with the Ce abundances of the \AGBCe~(similar results can be drawn for \AGBNd). The oxygen-rich nature of these sample stars is in agreement with model predictions although that, for metallicities around -0.70~dex and masses around 1.50~\Ms, FRUITY could predict a too-high surface carbon enrichment.
More importantly, the predicted and \gspspec\ Ce abundances are fully compatible
for AGB stars having an initial mass between $\sim$1.5 and $\sim$2.5~\Ms\ and metallicities between $\sim$-0.5 and $\sim$0.0~dex.

Finally, all these results confirm the excellent quality of the \Gaia\ data and of the \gspspec~chemical abundances, leading to a very large sample of AGB stars with second peak $s$-process elements abundances. Future studies may explore the abundances of other chemical species such as Sr, Y, Zr, Rb and/or Pb to better constrain the yield predictions.
\begin{acknowledgements}
      ARB and PdL acknowledge support from the European Union’s Horizon 2020 research and innovation program under SPACE-H2020 grant agreement number 101004214 (EXPLORE project).  PAP thanks the Centre National d'Etudes Spatiales (CNES) for funding support. C.A. acknowledges partial support by project PID2021-123110NB-I00 financed by MCIN/AEI /10.13039/501100011033/FEDER, UE. We sincerely thank Sergio Cristallo and Amanda Karakas for fruitful discussion. Special thanks to Niels Nieuwmunster for his grateful comments on some figures. \\
      This work has made use of data from the European Space Agency (ESA) mission \Gaia\ (https://www.cosmos.esa.int/gaia), processed by the \Gaia\ Data Processing and Analysis Consortium (DPAC, https://www.cosmos.esa.int/web/gaia/dpac/consortium). Funding for the DPAC has been provided by national institutions, in particular the institutions participating in the Gaia Multilateral Agreement. 
\end{acknowledgements}

%
%

\bibliographystyle{aa} 
\bibliography{ref}

\begin{thebibliography}{97}
\expandafter\ifx\csname natexlab\endcsname\relax\def\natexlab#1{#1}\fi

\bibitem[{{Abia} {et~al.}(2001){Abia}, {Busso}, {Gallino}, {Dom{\'\i}nguez},
  {Straniero}, \& {Isern}}]{Abia01}
{Abia}, C., {Busso}, M., {Gallino}, R., {et~al.} 2001, \apj, 559, 1117

\bibitem[{{Abia} {et~al.}(2022){Abia}, {de Laverny}, {Romero-G{\'o}mez}, \&
  {Figueras}}]{Abia22}
{Abia}, C., {de Laverny}, P., {Romero-G{\'o}mez}, M., \& {Figueras}, F. 2022,
  \aap, 664, A45

\bibitem[{{Abia} {et~al.}(2002){Abia}, {Dom{\'\i}nguez}, {Gallino}, {Busso},
  {Masera}, {Straniero}, {de Laverny}, {Plez}, \& {Isern}}]{Carlos02}
{Abia}, C., {Dom{\'\i}nguez}, I., {Gallino}, R., {et~al.} 2002, \apj, 579, 817

\bibitem[{{Arlandini} {et~al.}(1999){Arlandini}, {K{\"a}ppeler}, {Wisshak},
  {Gallino}, {Lugaro}, {Busso}, \& {Straniero}}]{Arlandini99}
{Arlandini}, C., {K{\"a}ppeler}, F., {Wisshak}, K., {et~al.} 1999, \apj, 525,
  886

\bibitem[{{Asplund} {et~al.}(2009){Asplund}, {Grevesse}, {Sauval}, \&
  {Scott}}]{Asplund09}
{Asplund}, M., {Grevesse}, N., {Sauval}, A.~J., \& {Scott}, P. 2009, \araa, 47,
  481

\bibitem[{{Bailer-Jones} {et~al.}(2021){Bailer-Jones}, {Rybizki}, {Fouesneau},
  {Demleitner}, \& {Andrae}}]{Coryn21}
{Bailer-Jones}, C.~A.~L., {Rybizki}, J., {Fouesneau}, M., {Demleitner}, M., \&
  {Andrae}, R. 2021, \aj, 161, 147

\bibitem[{{Baumgardt} \& {Hilker}(2018)}]{Bau18}
{Baumgardt}, H. \& {Hilker}, M. 2018, \mnras, 478, 1520

\bibitem[{{Belokurov} {et~al.}(2018){Belokurov}, {Erkal}, {Evans}, {Koposov},
  \& {Deason}}]{Belokurov18}
{Belokurov}, V., {Erkal}, D., {Evans}, N.~W., {Koposov}, S.~E., \& {Deason},
  A.~J. 2018, \mnras, 478, 611

\bibitem[{{Bijaoui}(2012)}]{2012ada..confE...2B}
{Bijaoui}, A. 2012, in Seventh Conference on Astronomical Data Analysis, ed.
  J.-L. {Starck} \& C.~{Surace}, 2

\bibitem[{{Birch} \& {Downs}(1994)}]{1994Metro..31..315B}
{Birch}, K.~P. \& {Downs}, M.~J. 1994, Metrologia, 31, 315

\bibitem[{{Bisterzo} {et~al.}(2015){Bisterzo}, {Gallino}, {K{\"a}ppeler},
  {Wiescher}, {Imbriani}, {Straniero}, {Cristallo}, {G{\"o}rres}, \&
  {deBoer}}]{Bisterzo15}
{Bisterzo}, S., {Gallino}, R., {K{\"a}ppeler}, F., {et~al.} 2015, \mnras, 449,
  506

\bibitem[{{Bisterzo} {et~al.}(2011){Bisterzo}, {Gallino}, {Straniero},
  {Cristallo}, \& {K{\"a}ppeler}}]{Bisterzo11}
{Bisterzo}, S., {Gallino}, R., {Straniero}, O., {Cristallo}, S., \&
  {K{\"a}ppeler}, F. 2011, \mnras, 418, 284

\bibitem[{{Bisterzo} {et~al.}(2016){Bisterzo}, {Travaglio}, {Wiescher},
  {Gallino}, {K{\"o}ppeler}, {Straniero}, {Cristallo}, {Imbriani},
  {G{\"o}rres}, \& {deBoer}}]{Bisterzo16}
{Bisterzo}, S., {Travaglio}, C., {Wiescher}, M., {et~al.} 2016, in Journal of
  Physics Conference Series, Vol. 665, Journal of Physics Conference Series,
  012023

\bibitem[{{Boesgaard} {et~al.}(2005){Boesgaard}, {King}, {Cody}, {Stephens}, \&
  {Deliyannis}}]{Boesgaard05}
{Boesgaard}, A.~M., {King}, J.~R., {Cody}, A.~M., {Stephens}, A., \&
  {Deliyannis}, C.~P. 2005, \apj, 629, 832

\bibitem[{{Busso} {et~al.}(2001){Busso}, {Gallino}, {Lambert}, {Travaglio}, \&
  {Smith}}]{Busso01}
{Busso}, M., {Gallino}, R., {Lambert}, D.~L., {Travaglio}, C., \& {Smith},
  V.~V. 2001, \apj, 557, 802

\bibitem[{{Busso} {et~al.}(1999){Busso}, {Gallino}, \& {Wasserburg}}]{Busso99}
{Busso}, M., {Gallino}, R., \& {Wasserburg}, G.~J. 1999, \araa, 37, 239

\bibitem[{{Cameron}(1960)}]{Cameron60}
{Cameron}, A.~G.~W. 1960, \aj, 65, 485

\bibitem[{{Carretta} {et~al.}(2009){Carretta}, {Bragaglia}, {Gratton},
  {D'Orazi}, \& {Lucatello}}]{Carretta09}
{Carretta}, E., {Bragaglia}, A., {Gratton}, R., {D'Orazi}, V., \& {Lucatello},
  S. 2009, \aap, 508, 695

\bibitem[{{Choplin} {et~al.}(2022){Choplin}, {Siess}, \& {Goriely}}]{Choplin22}
{Choplin}, A., {Siess}, L., \& {Goriely}, S. 2022, \aap, 667, A155

\bibitem[{{Contursi} {et~al.}(2021){Contursi}, {de Laverny}, {Recio-Blanco}, \&
  {Palicio}}]{BestArticleEver}
{Contursi}, G., {de Laverny}, P., {Recio-Blanco}, A., \& {Palicio}, P.~A. 2021,
  \aap, 654, A130

\bibitem[{{Contursi} {et~al.}(2023){Contursi}, {de Laverny}, {Recio-Blanco},
  {Spitoni}, {Palicio}, {Poggio}, {Grisoni}, {Cescutti}, {Matteucci}, {Spina},
  {{\'A}lvarez}, {Kordopatis}, {Ordenovic}, {Oreshina-Slezak}, \&
  {Zhao}}]{GOAT23}
{Contursi}, G., {de Laverny}, P., {Recio-Blanco}, A., {et~al.} 2023, \aap, 670,
  A106

\bibitem[{{Cristallo} {et~al.}(2016){Cristallo}, {Karinkuzhi}, {Goswami},
  {Piersanti}, \& {Gobrecht}}]{Cristallo16}
{Cristallo}, S., {Karinkuzhi}, D., {Goswami}, A., {Piersanti}, L., \&
  {Gobrecht}, D. 2016, \apj, 833, 181

\bibitem[{{Cristallo} {et~al.}(2011){Cristallo}, {Piersanti}, {Straniero},
  {Gallino}, {Dom{\'\i}nguez}, {Abia}, {Di Rico}, {Quintini}, \&
  {Bisterzo}}]{Cristallo11}
{Cristallo}, S., {Piersanti}, L., {Straniero}, O., {et~al.} 2011, \apjs, 197,
  17

\bibitem[{{Cristallo} {et~al.}(2009){Cristallo}, {Straniero}, {Gallino},
  {Piersanti}, {Dom{\'\i}nguez}, \& {Lederer}}]{Cristallo09}
{Cristallo}, S., {Straniero}, O., {Gallino}, R., {et~al.} 2009, \apj, 696, 797

\bibitem[{{Cristallo} {et~al.}(2015){Cristallo}, {Straniero}, {Piersanti}, \&
  {Gobrecht}}]{Cristallo15}
{Cristallo}, S., {Straniero}, O., {Piersanti}, L., \& {Gobrecht}, D. 2015,
  \apjs, 219, 40

\bibitem[{{Cseh} {et~al.}(2022){Cseh}, {Vil{\'a}gos}, {Roriz}, {Pereira},
  {D'Orazi}, {Karakas}, {So{\'o}s}, {Drake}, {Junqueira}, \& {Lugaro}}]{Cseh22}
{Cseh}, B., {Vil{\'a}gos}, B., {Roriz}, M.~P., {et~al.} 2022, \aap, 660, A128

\bibitem[{{Danziger}(1966)}]{Danziger66}
{Danziger}, I.~J. 1966, \apj, 143, 527

\bibitem[{{Den Hartog} {et~al.}(2003){Den Hartog}, {Lawler}, {Sneden}, \&
  {Cowan}}]{DenHartog2003}
{Den Hartog}, E.~A., {Lawler}, J.~E., {Sneden}, C., \& {Cowan}, J.~J. 2003,
  \apjs, 148, 543

\bibitem[{{Fanelli} {et~al.}(2021){Fanelli}, {Origlia}, {Oliva}, {Mucciarelli},
  {Sanna}, {Dalessandro}, \& {Romano}}]{Fanelli21}
{Fanelli}, C., {Origlia}, L., {Oliva}, E., {et~al.} 2021, \aap, 645, A19

\bibitem[{{Feuillet} {et~al.}(2020){Feuillet}, {Feltzing}, {Sahlholdt}, \&
  {Casagrande}}]{Feuillet20}
{Feuillet}, D.~K., {Feltzing}, S., {Sahlholdt}, C.~L., \& {Casagrande}, L.
  2020, \mnras, 497, 109

\bibitem[{{Feuillet} {et~al.}(2021){Feuillet}, {Sahlholdt}, {Feltzing}, \&
  {Casagrande}}]{Feuillet21}
{Feuillet}, D.~K., {Sahlholdt}, C.~L., {Feltzing}, S., \& {Casagrande}, L.
  2021, \mnras, 508, 1489

\bibitem[{{Fishlock} {et~al.}(2014){Fishlock}, {Karakas}, {Lugaro}, \&
  {Yong}}]{Fishlock14}
{Fishlock}, C.~K., {Karakas}, A.~I., {Lugaro}, M., \& {Yong}, D. 2014, \apj,
  797, 44

\bibitem[{{Gaia Collaboration} {et~al.}(2023){Gaia Collaboration},
  {Recio-Blanco}, {Kordopatis}, {de Laverny}, {Palicio}, {Spagna}, {Spina},
  {Katz}, {Re Fiorentin}, {Poggio}, {McMillan}, {Vallenari}, {Lattanzi},
  {Seabroke}, {Casamiquela}, {Bragaglia}, {Antoja}, {Bailer-Jones},
  {Schultheis}, {Andrae}, {Fouesneau}, {Cropper}, {Cantat-Gaudin}, {Bijaoui},
  {Heiter}, {Brown}, {Prusti}, {de Bruijne}, {Arenou}, {Babusiaux}, {Biermann},
  {Creevey}, {Ducourant}, {Evans}, {Eyer}, {Guerra}, {Hutton}, {Jordi},
  {Klioner}, {Lammers}, {Lindegren}, {Luri}, {Mignard}, {Panem}, {Pourbaix},
  {Randich}, {Sartoretti}, {Soubiran}, {Tanga}, {Walton}, {Bastian}, {Drimmel},
  {Jansen}, {van Leeuwen}, {Bakker}, {Cacciari}, {Casta{\~n}eda}, {De Angeli},
  {Fabricius}, {Fr{\'e}mat}, {Galluccio}, {Guerrier}, {Masana}, {Messineo},
  {Mowlavi}, {Nicolas}, {Nienartowicz}, {Pailler}, {Panuzzo}, {Riclet}, {Roux},
  {Sordo}, {Th{\'e}venin}, {Gracia-Abril}, {Portell}, {Teyssier}, {Altmann},
  {Audard}, {Bellas-Velidis}, {Benson}, {Berthier}, {Blomme}, {Burgess},
  {Busonero}, {Busso}, {C{\'a}novas}, {Carry}, {Cellino}, {Cheek},
  {Clementini}, {Damerdji}, {Davidson}, {de Teodoro}, {Nu{\~n}ez Campos},
  {Delchambre}, {Dell'Oro}, {Esquej}, {Fern{\'a}ndez-Hern{\'a}ndez}, {Fraile},
  {Garabato}, {Garc{\'\i}a-Lario}, {Gosset}, {Haigron}, {Halbwachs}, {Hambly},
  {Harrison}, {Hern{\'a}ndez}, {Hestroffer}, {Hodgkin}, {Holl}, {Jan{\ss}en},
  {Jevardat de Fombelle}, {Jordan}, {Krone-Martins}, {Lanzafame},
  {L{\"o}ffler}, {Marchal}, {Marrese}, {Moitinho}, {Muinonen}, {Osborne},
  {Pancino}, {Pauwels}, {Reyl{\'e}}, {Riello}, {Rimoldini}, {Roegiers},
  {Rybizki}, {Sarro}, {Siopis}, {Smith}, {Sozzetti}, {Utrilla}, {van Leeuwen},
  {Abbas}, {{\'A}brah{\'a}m}, {Abreu Aramburu}, {Aerts}, {Aguado}, {Ajaj},
  {Aldea-Montero}, {Altavilla}, {{\'A}lvarez}, {Alves}, {Anders}, {Anderson},
  {Anglada Varela}, {Baines}, {Baker}, {Balaguer-N{\'u}{\~n}ez}, {Balbinot},
  {Balog}, {Barache}, {Barbato}, {Barros}, {Barstow}, {Bartolom{\'e}},
  {Bassilana}, {Bauchet}, {Becciani}, {Bellazzini}, {Berihuete}, {Bernet},
  {Bertone}, {Bianchi}, {Binnenfeld}, {Blanco-Cuaresma}, {Boch}, {Bombrun},
  {Bossini}, {Bouquillon}, {Bramante}, {Breedt}, {Bressan}, {Brouillet},
  {Brugaletta}, {Bucciarelli}, {Burlacu}, {Butkevich}, {Buzzi}, {Caffau},
  {Cancelliere}, {Carballo}, {Carlucci}, {Carnerero}, {Carrasco}, {Castellani},
  {Castro-Ginard}, {Chaoul}, {Charlot}, {Chemin}, {Chiaramida}, {Chiavassa},
  {Chornay}, {Comoretto}, {Contursi}, {Cooper}, {Cornez}, {Cowell}, {Crifo},
  {Crosta}, {Crowley}, {Dafonte}, {Dapergolas}, {David}, {De Luise}, {De
  March}, {De Ridder}, {de Souza}, {de Torres}, {del Peloso}, {del Pozo},
  {Delbo}, {Delgado}, {Delisle}, {Demouchy}, {Dharmawardena}, {Di Matteo},
  {Diakite}, {Diener}, {Distefano}, {Dolding}, {Edvardsson}, {Enke}, {Fabre},
  {Fabrizio}, {Faigler}, {Fedorets}, {Fernique}, {Figueras}, {Fournier},
  {Fouron}, {Fragkoudi}, {Gai}, {Garcia-Gutierrez}, {Garcia-Reinaldos},
  {Garc{\'\i}a-Torres}, {Garofalo}, {Gavel}, {Gavras}, {Gerlach}, {Geyer},
  {Giacobbe}, {Gilmore}, {Girona}, {Giuffrida}, {Gomel}, {Gomez},
  {Gonz{\'a}lez-N{\'u}{\~n}ez}, {Gonz{\'a}lez-Santamar{\'\i}a},
  {Gonz{\'a}lez-Vidal}, {Granvik}, {Guillout}, {Guiraud},
  {Guti{\'e}rrez-S{\'a}nchez}, {Guy}, {Hatzidimitriou}, {Hauser}, {Haywood},
  {Helmer}, {Helmi}, {Sarmiento}, {Hidalgo}, {H{\l}adczuk}, {Hobbs}, {Holland},
  {Huckle}, {Jardine}, {Jasniewicz}, {Jean-Antoine Piccolo},
  {Jim{\'e}nez-Arranz}, {Juaristi Campillo}, {Julbe}, {Karbevska}, {Kervella},
  {Khanna}, {Korn}, {K{\'o}sp{\'a}l}, {Kostrzewa-Rutkowska}, {Kruszy{\'n}ska},
  {Kun}, {Laizeau}, {Lambert}, {Lanza}, {Lasne}, {Le Campion}, {Lebreton},
  {Lebzelter}, {Leccia}, {Leclerc}, {Lecoeur-Taibi}, {Liao}, {Licata},
  {Lindstr{\o}m}, {Lister}, {Livanou}, {Lobel}, {Lorca}, {Loup}, {Madrero
  Pardo}, {Magdaleno Romeo}, {Managau}, {Mann}, {Manteiga}, {Marchant},
  {Marconi}, {Marcos}, {Marcos Santos}, {Mar{\'\i}n Pina}, {Marinoni},
  {Marocco}, {Marshall}, {Martin Polo}, {Mart{\'\i}n-Fleitas}, {Marton},
  {Mary}, {Masip}, {Massari}, {Mastrobuono-Battisti}, {Mazeh}, {Messina},
  {Michalik}, {Millar}, {Mints}, {Molina}, {Molinaro}, {Moln{\'a}r}, {Monari},
  {Mongui{\'o}}, {Montegriffo}, {Montero}, {Mor}, {Mora}, {Morbidelli},
  {Morel}, {Morris}, {Muraveva}, {Murphy}, {Musella}, {Nagy}, {Noval},
  {Oca{\~n}a}, {Ogden}, {Ordenovic}, {Osinde}, {Pagani}, {Pagano}, {Palaversa},
  {Pallas-Quintela}, {Panahi}, {Payne-Wardenaar}, {Pe{\~n}alosa Esteller},
  {Penttil{\"a}}, {Pichon}, {Piersimoni}, {Pineau}, {Plachy}, {Plum},
  {Pr{\v{s}}a}, {Pulone}, {Racero}, {Ragaini}, {Rainer}, {Raiteri}, {Ramos},
  {Ramos-Lerate}, {Regibo}, {Richards}, {Rios Diaz}, {Ripepi}, {Riva}, {Rix},
  {Rixon}, {Robichon}, {Robin}, {Robin}, {Roelens}, {Rogues}, {Rohrbasser},
  {Romero-G{\'o}mez}, {Rowell}, {Royer}, {Ruz Mieres}, {Rybicki}, {Sadowski},
  {S{\'a}ez N{\'u}{\~n}ez}, {Sagrist{\`a} Sell{\'e}s}, {Sahlmann}, {Salguero},
  {Samaras}, {Sanchez Gimenez}, {Sanna}, {Santove{\~n}a}, {Sarasso}, {Sciacca},
  {Segol}, {Segovia}, {S{\'e}gransan}, {Semeux}, {Shahaf}, {Siddiqui},
  {Siebert}, {Siltala}, {Silvelo}, {Slezak}, {Slezak}, {Smart}, {Snaith},
  {Solano}, {Solitro}, {Souami}, {Souchay}, {Spoto}, {Steele},
  {Steidelm{\"u}ller}, {Stephenson}, {S{\"u}veges}, {Surdej}, {Szabados},
  {Szegedi-Elek}, {Taris}, {Taylor}, {Teixeira}, {Tolomei}, {Tonello}, {Torra},
  {Torra}, {Torralba Elipe}, {Trabucchi}, {Tsounis}, {Turon}, {Ulla}, {Unger},
  {Vaillant}, {van Dillen}, {van Reeven}, {Vanel}, {Vecchiato}, {Viala},
  {Vicente}, {Voutsinas}, {Weiler}, {Wevers}, {Wyrzykowski}, {Yoldas}, {Yvard},
  {Zhao}, {Zorec}, {Zucker}, \& {Zwitter}}]{PVP_Ale}
{Gaia Collaboration}, {Recio-Blanco}, A., {Kordopatis}, G., {et~al.} 2023,
  \aap, 674, A38

\bibitem[{{Gaia Collaboration, Vallenari} {et~al.}(2022){Gaia Collaboration,
  Vallenari}, {A.}, {Brown, A.G.A.}, {Prusti, T.}, \& {et al.}}]{Vallenari22}
{Gaia Collaboration, Vallenari}, {A.}, {Brown, A.G.A.}, {Prusti, T.}, \& {et
  al.} 2022, A\&A

\bibitem[{{Gallino} {et~al.}(1998){Gallino}, {Arlandini}, {Busso}, {Lugaro},
  {Travaglio}, {Straniero}, {Chieffi}, \& {Limongi}}]{Gallino98}
{Gallino}, R., {Arlandini}, C., {Busso}, M., {et~al.} 1998, \apj, 497, 388

\bibitem[{{Gerber} {et~al.}(2020){Gerber}, {Friel}, \& {Vesperini}}]{Greber20}
{Gerber}, J.~M., {Friel}, E.~D., \& {Vesperini}, E. 2020, \aj, 159, 50

\bibitem[{{Goriely} \& {Siess}(2004)}]{Goriely-Siess04}
{Goriely}, S. \& {Siess}, L. 2004, \aap, 421, L25

\bibitem[{{Grevesse} {et~al.}(2007){Grevesse}, {Asplund}, \&
  {Sauval}}]{Grevesse07}
{Grevesse}, N., {Asplund}, M., \& {Sauval}, A.~J. 2007, \ssr, 130, 105

\bibitem[{{Harris}(1996)}]{Harris96}
{Harris}, W.~E. 1996, \aj, 112, 1487

\bibitem[{{Hayes} {et~al.}(2022){Hayes}, {Masseron}, {Sobeck},
  {Garc{\'\i}a-Hern{\'a}ndez}, {Prieto}, {Beaton}, {Cunha}, {Hasselquist},
  {Holtzman}, {J{\"o}nsson}, {Majewski}, {Shetrone}, {Smith}, \&
  {Almeida}}]{BAWLAS}
{Hayes}, C.~R., {Masseron}, T., {Sobeck}, J., {et~al.} 2022, \apjs, 262, 34

\bibitem[{{Helmi}(2020)}]{Helmi20}
{Helmi}, A. 2020, \araa, 58, 205

\bibitem[{{Helmi} {et~al.}(2018){Helmi}, {Babusiaux}, {Koppelman}, {Massari},
  {Veljanoski}, \& {Brown}}]{Helmi18}
{Helmi}, A., {Babusiaux}, C., {Koppelman}, H.~H., {et~al.} 2018, \nat, 563, 85

\bibitem[{{Herwig}(2004)}]{Herwig04}
{Herwig}, F. 2004, \apj, 605, 425

\bibitem[{{Hinkle} {et~al.}(2000){Hinkle}, {Wallace}, {Valenti}, \&
  {Harmer}}]{Hinkle00}
{Hinkle}, K., {Wallace}, L., {Valenti}, J., \& {Harmer}, D. 2000, {Visible and
  Near Infrared Atlas of the Arcturus Spectrum 3727-9300 A}

\bibitem[{{Iben}(1975)}]{Iben75}
{Iben}, I., J. 1975, \apj, 196, 549

\bibitem[{{Iben} \& {Renzini}(1983)}]{IR83}
{Iben}, I., J. \& {Renzini}, A. 1983, \araa, 21, 271

\bibitem[{{J{\"o}nsson} {et~al.}(2020){J{\"o}nsson}, {Holtzman}, {Allende
  Prieto}, {Cunha}, {Garc{\'\i}a-Hern{\'a}ndez}, {Hasselquist}, {Masseron},
  {Osorio}, {Shetrone}, {Smith}, {Stringfellow}, {Bizyaev}, {Edvardsson},
  {Majewski}, {M{\'e}sz{\'a}ros}, {Souto}, {Zamora}, {Beaton}, {Bovy}, {Donor},
  {Pinsonneault}, {Poovelil}, \& {Sobeck}}]{Jonsson20}
{J{\"o}nsson}, H., {Holtzman}, J.~A., {Allende Prieto}, C., {et~al.} 2020, \aj,
  160, 120

\bibitem[{{K{\"a}ppeler} {et~al.}(2011){K{\"a}ppeler}, {Gallino}, {Bisterzo},
  \& {Aoki}}]{Kappeler11}
{K{\"a}ppeler}, F., {Gallino}, R., {Bisterzo}, S., \& {Aoki}, W. 2011, Reviews
  of Modern Physics, 83, 157

\bibitem[{{Karakas} \& {Lattanzio}(2014{\natexlab{a}})}]{KL14}
{Karakas}, A.~I. \& {Lattanzio}, J.~C. 2014{\natexlab{a}}, \pasa, 31, e030

\bibitem[{{Karakas} \& {Lattanzio}(2014{\natexlab{b}})}]{Karakas14}
{Karakas}, A.~I. \& {Lattanzio}, J.~C. 2014{\natexlab{b}}, \pasa, 31, e030

\bibitem[{{Karakas} {et~al.}(2002){Karakas}, {Lattanzio}, \&
  {Pols}}]{Karakas02}
{Karakas}, A.~I., {Lattanzio}, J.~C., \& {Pols}, O.~R. 2002, \pasa, 19, 515

\bibitem[{{Karakas} \& {Lugaro}(2016)}]{Karakas16}
{Karakas}, A.~I. \& {Lugaro}, M. 2016, \apj, 825, 26

\bibitem[{{Karakas} {et~al.}(2018){Karakas}, {Lugaro}, {Carlos}, {Cseh},
  {Kamath}, \& {Garc{\'\i}a-Hern{\'a}ndez}}]{Karakas18}
{Karakas}, A.~I., {Lugaro}, M., {Carlos}, M., {et~al.} 2018, \mnras, 477, 421

\bibitem[{{Koppelman} {et~al.}(2019){Koppelman}, {Helmi}, {Massari},
  {Price-Whelan}, \& {Starkenburg}}]{Koppelman19}
{Koppelman}, H.~H., {Helmi}, A., {Massari}, D., {Price-Whelan}, A.~M., \&
  {Starkenburg}, T.~K. 2019, \aap, 631, L9

\bibitem[{{Lamb} {et~al.}(1977){Lamb}, {Howard}, {Truran}, \& {Iben}}]{Lamb77}
{Lamb}, S.~A., {Howard}, W.~M., {Truran}, J.~W., \& {Iben}, I., J. 1977, \apj,
  217, 213

\bibitem[{{Lambert}(1991)}]{Lambert91}
{Lambert}, D.~L. 1991, in Evolution of Stars: the Photospheric Abundance
  Connection, ed. G.~{Michaud} \& A.~V. {Tutukov}, Vol. 145, 299

\bibitem[{{Lambert} {et~al.}(1995){Lambert}, {Smith}, {Busso}, {Gallino}, \&
  {Straniero}}]{Lambert95}
{Lambert}, D.~L., {Smith}, V.~V., {Busso}, M., {Gallino}, R., \& {Straniero},
  O. 1995, \apj, 450, 302

\bibitem[{{Lebzelter} {et~al.}(2023){Lebzelter}, {Mowlavi}, {Lecoeur-Taibi},
  {Trabucchi}, {Audard}, {Garc{\'\i}a-Lario}, {Gavras}, {Holl}, {Jevardat de
  Fombelle}, {Nienartowicz}, {Rimoldini}, \& {Eyer}}]{Leb22}
{Lebzelter}, T., {Mowlavi}, N., {Lecoeur-Taibi}, I., {et~al.} 2023, \aap, 674,
  A15

\bibitem[{{Lebzelter} {et~al.}(2018){Lebzelter}, {Mowlavi}, {Marigo},
  {Pastorelli}, {Trabucchi}, {Wood}, \& {Lecoeur-Ta{\"\i}bi}}]{Lebzelter18}
{Lebzelter}, T., {Mowlavi}, N., {Marigo}, P., {et~al.} 2018, \aap, 616, L13

\bibitem[{{Limongi} \& {Chieffi}(2018)}]{Limongi18}
{Limongi}, M. \& {Chieffi}, A. 2018, \apjs, 237, 13

\bibitem[{{Lodders}(2003)}]{Lodders03}
{Lodders}, K. 2003, \apj, 591, 1220

\bibitem[{{Lugaro} {et~al.}(2016){Lugaro}, {Campbell}, {D'Orazi}, {Karakas},
  {Garcia-Hernandez}, {Stancliffe}, {Tagliente}, {Iliadis}, \&
  {Rauscher}}]{Lugaro16}
{Lugaro}, M., {Campbell}, S.~W., {D'Orazi}, V., {et~al.} 2016, in Journal of
  Physics Conference Series, Vol. 665, Journal of Physics Conference Series,
  012021

\bibitem[{{Lugaro} {et~al.}(2012){Lugaro}, {Karakas}, {Stancliffe}, \&
  {Rijs}}]{Lugaro12}
{Lugaro}, M., {Karakas}, A.~I., {Stancliffe}, R.~J., \& {Rijs}, C. 2012, \apj,
  747, 2

\bibitem[{{Marigo}(2002)}]{Marigo02}
{Marigo}, P. 2002, \aap, 387, 507

\bibitem[{{Masseron} {et~al.}(2019){Masseron}, {Garc{\'\i}a-Hern{\'a}ndez},
  {M{\'e}sz{\'a}ros}, {Zamora}, {Dell'Agli}, {Allende Prieto}, {Edvardsson},
  {Shetrone}, {Plez}, {Fern{\'a}ndez-Trincado}, {Cunha}, {J{\"o}nsson},
  {Geisler}, {Beers}, \& {Cohen}}]{Masseron19}
{Masseron}, T., {Garc{\'\i}a-Hern{\'a}ndez}, D.~A., {M{\'e}sz{\'a}ros}, S.,
  {et~al.} 2019, \aap, 622, A191

\bibitem[{{Matsuno} {et~al.}(2022){Matsuno}, {Koppelman}, {Helmi}, {Aoki},
  {Ishigaki}, {Suda}, {Yuan}, \& {Hattori}}]{Matsuno22}
{Matsuno}, T., {Koppelman}, H.~H., {Helmi}, A., {et~al.} 2022, \aap, 661, A103

\bibitem[{{Messineo}(2023)}]{Messineo23}
{Messineo}, M. 2023, \aap, 671, A148

\bibitem[{{Myeong} {et~al.}(2018){Myeong}, {Evans}, {Belokurov}, {Amorisco}, \&
  {Koposov}}]{Myeong18}
{Myeong}, G.~C., {Evans}, N.~W., {Belokurov}, V., {Amorisco}, N.~C., \&
  {Koposov}, S.~E. 2018, \mnras, 475, 1537

\bibitem[{{Myeong} {et~al.}(2019){Myeong}, {Vasiliev}, {Iorio}, {Evans}, \&
  {Belokurov}}]{Myeong19}
{Myeong}, G.~C., {Vasiliev}, E., {Iorio}, G., {Evans}, N.~W., \& {Belokurov},
  V. 2019, \mnras, 488, 1235

\bibitem[{{Palicio} {et~al.}(2023){Palicio}, {Recio-Blanco}, {Poggio},
  {Antoja}, {McMillan}, \& {Spitoni}}]{Pedro23}
{Palicio}, P.~A., {Recio-Blanco}, A., {Poggio}, E., {et~al.} 2023, \aap, 670,
  L7

\bibitem[{{Peters}(1968)}]{Peters68}
{Peters}, J.~G. 1968, \apj, 154, 225

\bibitem[{{Pignatari} {et~al.}(2010){Pignatari}, {Gallino}, {Heil}, {Wiescher},
  {K{\"a}ppeler}, {Herwig}, \& {Bisterzo}}]{Pignatari10}
{Pignatari}, M., {Gallino}, R., {Heil}, M., {et~al.} 2010, \apj, 710, 1557

\bibitem[{{Prantzos} {et~al.}(2020){Prantzos}, {Abia}, {Cristallo}, {Limongi},
  \& {Chieffi}}]{Prantzos20}
{Prantzos}, N., {Abia}, C., {Cristallo}, S., {Limongi}, M., \& {Chieffi}, A.
  2020, \mnras, 491, 1832

\bibitem[{{Prantzos} {et~al.}(2018){Prantzos}, {Abia}, {Limongi}, {Chieffi}, \&
  {Cristallo}}]{Prantzos18}
{Prantzos}, N., {Abia}, C., {Limongi}, M., {Chieffi}, A., \& {Cristallo}, S.
  2018, \mnras, 476, 3432

\bibitem[{{Ram{\'\i}rez} \& {Cohen}(2002)}]{Ramirez02}
{Ram{\'\i}rez}, S.~V. \& {Cohen}, J.~G. 2002, \aj, 123, 3277

\bibitem[{{Recio-Blanco} {et~al.}(2016){Recio-Blanco}, {de Laverny}, {Allende
  Prieto}, {Fustes}, {Manteiga}, {Arcay}, {Bijaoui}, {Dafonte}, {Ordenovic}, \&
  {Ordo{\~n}ez Blanco}}]{RB16}
{Recio-Blanco}, A., {de Laverny}, P., {Allende Prieto}, C., {et~al.} 2016,
  \aap, 585, A93

\bibitem[{{Recio-Blanco} {et~al.}(2023){Recio-Blanco}, {de Laverny}, {Palicio},
  {Kordopatis}, {{\'A}lvarez}, {Schultheis}, {Contursi}, {Zhao}, {Torralba
  Elipe}, {Ordenovic}, {Manteiga}, {Dafonte}, {Oreshina-Slezak}, {Bijaoui},
  {Fr{\'e}mat}, {Seabroke}, {Pailler}, {Spitoni}, {Poggio}, {Creevey}, {Abreu
  Aramburu}, {Accart}, {Andrae}, {Bailer-Jones}, {Bellas-Velidis}, {Brouillet},
  {Brugaletta}, {Burlacu}, {Carballo}, {Casamiquela}, {Chiavassa}, {Cooper},
  {Dapergolas}, {Delchambre}, {Dharmawardena}, {Drimmel}, {Edvardsson},
  {Fouesneau}, {Garabato}, {Garc{\'\i}a-Lario}, {Garc{\'\i}a-Torres}, {Gavel},
  {Gomez}, {Gonz{\'a}lez-Santamar{\'\i}a}, {Hatzidimitriou}, {Heiter},
  {Jean-Antoine Piccolo}, {Kontizas}, {Korn}, {Lanzafame}, {Lebreton}, {Le
  Fustec}, {Licata}, {Lindstr{\o}m}, {Livanou}, {Lobel}, {Lorca}, {Magdaleno
  Romeo}, {Marocco}, {Marshall}, {Mary}, {Nicolas}, {Pallas-Quintela}, {Panem},
  {Pichon}, {Riclet}, {Robin}, {Rybizki}, {Santove{\~n}a}, {Silvelo}, {Smart},
  {Sarro}, {Sordo}, {Soubiran}, {S{\"u}veges}, {Ulla}, {Vallenari}, {Zorec},
  {Utrilla}, \& {Bakker}}]{GSPspecDR3}
{Recio-Blanco}, A., {de Laverny}, P., {Palicio}, P.~A., {et~al.} 2023, \aap,
  674, A29

\bibitem[{{Recio-Blanco} {et~al.}(2021){Recio-Blanco}, {Fern{\'a}ndez-Alvar},
  {de Laverny}, {Antoja}, {Helmi}, \& {Crida}}]{RecioBlanco2021}
{Recio-Blanco}, A., {Fern{\'a}ndez-Alvar}, E., {de Laverny}, P., {et~al.} 2021,
  \aap, 648, A108

\bibitem[{{Roriz} {et~al.}(2021){Roriz}, {Lugaro}, {Pereira}, {Drake},
  {Junqueira}, \& {Sneden}}]{Roriz21}
{Roriz}, M.~P., {Lugaro}, M., {Pereira}, C.~B., {et~al.} 2021, \mnras, 501,
  5834

\bibitem[{{Rybizki} {et~al.}(2022){Rybizki}, {Green}, {Rix}, {El-Badry},
  {Demleitner}, {Zari}, {Udalski}, {Smart}, \& {Gould}}]{Rybizki22}
{Rybizki}, J., {Green}, G.~M., {Rix}, H.-W., {et~al.} 2022, \mnras, 510, 2597

\bibitem[{{Sanders} \& {Matsunaga}(2023)}]{Sanders23}
{Sanders}, J.~L. \& {Matsunaga}, N. 2023, \mnras, 521, 2745

\bibitem[{{Shejeelammal} {et~al.}(2020){Shejeelammal}, {Goswami}, {Goswami},
  {Rathour}, \& {Masseron}}]{Shejeelammal20}
{Shejeelammal}, J., {Goswami}, A., {Goswami}, P.~P., {Rathour}, R.~S., \&
  {Masseron}, T. 2020, \mnras, 492, 3708

\bibitem[{{Skrutskie} {et~al.}(2006){Skrutskie}, {Cutri}, {Stiening},
  {Weinberg}, {Schneider}, {Carpenter}, {Beichman}, {Capps}, {Chester},
  {Elias}, {Huchra}, {Liebert}, {Lonsdale}, {Monet}, {Price}, {Seitzer},
  {Jarrett}, {Kirkpatrick}, {Gizis}, {Howard}, {Evans}, {Fowler}, {Fullmer},
  {Hurt}, {Light}, {Kopan}, {Marsh}, {McCallon}, {Tam}, {Van Dyk}, \&
  {Wheelock}}]{Skrutskie06}
{Skrutskie}, M.~F., {Cutri}, R.~M., {Stiening}, R., {et~al.} 2006, \aj, 131,
  1163

\bibitem[{{Smith} \& {Lambert}(1985)}]{Smith85}
{Smith}, V.~V. \& {Lambert}, D.~L. 1985, \apj, 294, 326

\bibitem[{{Smith} \& {Lambert}(1986)}]{Smith86}
{Smith}, V.~V. \& {Lambert}, D.~L. 1986, \apj, 311, 843

\bibitem[{{Smith} \& {Lambert}(1988)}]{Smith88}
{Smith}, V.~V. \& {Lambert}, D.~L. 1988, \apj, 333, 219

\bibitem[{{Smith} {et~al.}(1987){Smith}, {Lambert}, \& {McWilliam}}]{Smith87}
{Smith}, V.~V., {Lambert}, D.~L., \& {McWilliam}, A. 1987, \apj, 320, 862

\bibitem[{{Sneden} {et~al.}(1994){Sneden}, {Kraft}, {Langer}, {Prosser}, \&
  {Shetrone}}]{Sneden94}
{Sneden}, C., {Kraft}, R.~P., {Langer}, G.~E., {Prosser}, C.~F., \& {Shetrone},
  M.~D. 1994, \aj, 107, 1773

\bibitem[{{Soszy{\'n}ski} {et~al.}(2005){Soszy{\'n}ski}, {Gieren}, \&
  {Pietrzy{\'n}ski}}]{So05}
{Soszy{\'n}ski}, I., {Gieren}, W., \& {Pietrzy{\'n}ski}, G. 2005, \pasp, 117,
  823

\bibitem[{{Straniero} {et~al.}(2023){Straniero}, {Abia}, \&
  {Dom{\'\i}nguez}}]{Straniero23}
{Straniero}, O., {Abia}, C., \& {Dom{\'\i}nguez}, I. 2023, European Physical
  Journal A, 59, 17

\bibitem[{{Straniero} {et~al.}(2014){Straniero}, {Cristallo}, \&
  {Piersanti}}]{Straniero14}
{Straniero}, O., {Cristallo}, S., \& {Piersanti}, L. 2014, \apj, 785, 77

\bibitem[{{Straniero} {et~al.}(2003){Straniero}, {Dom{\'\i}nguez}, {Cristallo},
  \& {Gallino}}]{Straniero03}
{Straniero}, O., {Dom{\'\i}nguez}, I., {Cristallo}, S., \& {Gallino}, R. 2003,
  \pasa, 20, 389

\bibitem[{{Straniero} {et~al.}(1995){Straniero}, {Gallino}, {Busso}, {Chiefei},
  {Raiteri}, {Limongi}, \& {Salaris}}]{Straniero95}
{Straniero}, O., {Gallino}, R., {Busso}, M., {et~al.} 1995, \apjl, 440, L85

\bibitem[{{Tautvai{\v{s}}ien{\.{e}}} {et~al.}(2021){Tautvai{\v{s}}ien{\.{e}}},
  {Viscasillas V{\'a}zquez}, {Mikolaitis}, {Stonkut{\.{e}}},
  {Minkevi{\v{c}}i{\={u}}t{\.{e}}}, {Drazdauskas}, \& {Bagdonas}}]{Taut21}
{Tautvai{\v{s}}ien{\.{e}}}, G., {Viscasillas V{\'a}zquez}, C., {Mikolaitis},
  {\v{S}}., {et~al.} 2021, \aap, 649, A126

\bibitem[{{Ulrich}(1973)}]{Ulrich73}
{Ulrich}, R.~K. 1973, in Explosive Nucleosynthesis, ed. D.~N. {Schramm} \&
  W.~D. {Arnett}, 139

\bibitem[{{Utsumi}(1970)}]{Utsumi70}
{Utsumi}, K. 1970, \pasj, 22, 93

\bibitem[{{Vanture}(1992)}]{Vanture92}
{Vanture}, A.~D. 1992, \aj, 104, 1997

\end{thebibliography}

\begin{appendix}
\section{ADQL query}
\label{Append}
\lstset{language=SQL}

\begin{lstlisting}[caption={\texttt{ADQL} query for the \AGBNd. Note that this query returns non-calibrated \g.},captionpos=b]
SELECT source_id
FROM gaiadr3.astrophysical_parameters inner join
gaiadr3.gaia_source using(source_id)
WHERE
(rv_expected_sig_to_noise>0)
AND
(vbroad<=13)
AND
(teff_gspspec IS NOT NULL)
AND
(teff_gspspec<=4000)
AND
(ndfe_gspspec IS NOT NULL)
AND
( (ndfe_gspspec_upper-ndfe_gspspec_lower)<=0.40)
AND
(flags_gspspec LIKE '0%')
AND
(flags_gspspec LIKE '_0%')
AND
(flags_gspspec LIKE '__0%')
AND
(flags_gspspec LIKE '___0%')
AND
(flags_gspspec LIKE '____0%')
AND
(flags_gspspec LIKE '_____0%')
AND
(flags_gspspec LIKE '______0%')
AND
((flags_gspspec LIKE '_______0%') OR (flags_gspspec
LIKE '_______1%') )
AND
(flags_gspspec LIKE '________0%')
AND
(flags_gspspec LIKE '_________0%')
AND
(flags_gspspec LIKE '__________0%')
AND
(flags_gspspec LIKE '___________0%')
AND
(flags_gspspec LIKE '____________0%')
AND
((flags_gspspec LIKE '%0___') OR (flags_gspspec
LIKE '%1___') OR (flags_gspspec LIKE '%2___'))
AND
((flags_gspspec LIKE '%0__') OR (flags_gspspec LIKE
'%1__') )
AND
NOT
((ndfe_gspspec>1.99) AND (flags_gspspec LIKE '____________0%') AND (flags_gspspec NOT LIKE '%0___'))
\end{lstlisting}
\end{appendix}

\end{document}